\documentclass[12pt,preprint]{aastex}

\def\d{{\rm d}}

\shorttitle{Flux and Spectral Index Correlation}

\begin{document}


\title{Correlation between Flux and Spectral Index  during Flares in 
Sagittarius A*}


\author{Jonathan M. Bittner,\altaffilmark{1} Siming Liu,\altaffilmark{2} 
Christopher L. Fryer,\altaffilmark{2, 3} and
Vah\'e Petrosian\altaffilmark{4} 
}

\altaffiltext{1}{Physics Department, Yale University, New Haven, CT 
06520-8120; jonathan.bittner@yale.edu}
\altaffiltext{2}{Los Alamos National Laboratory, Los Alamos, NM 87545; 
liusm@lanl.gov}
\altaffiltext{3}{Physics Department, The University of Arizona, Tucson, AZ 
85721; fryer@lanl.gov}
\altaffiltext{4}{Center for Space Science and Astrophysics, Department of 
Physics and Applied Physics, Stanford 
University, Stanford, CA 94305; vahe@astronomy.stanford.edu}


\begin{abstract}
Flares in Sagittarius A* are produced by hot plasmas within a few Schwarzschild radii of the supermassive black hole at the 
Galactic center. The recent detection of a correlation between the spectral index and flux during a near infrared (NIR) flare 
provides a means to conduct detailed investigations of the plasma heating and radiation processes. We study the evolution of 
the electron distribution function under the influence of a turbulent magnetic field in a hot collisionless plasma.  The 
magnetic field, presumably generated through instabilities in the accretion flow, can both heat the plasma via 
resonant wave-particle coupling and cool the electrons via radiation.  The electron distribution can generally be 
approximated as relativistic Maxwellian. To account for the observed correlation, we find that the magnetic field needs 
to be anti-correlated with the electron ``temperature''. NIR and X-ray light curves are produced for a cooling and a heating 
phase. The model predicts simultaneous flare activity in the NIR and X-ray bands, which can be compared with 
observations. These results can be applied to MHD simulations to study the radiative characteristics of collisionless 
plasmas, especially accretion flows in low-luminosity AGNs.
\end{abstract}





\section{Introduction}

It is generally accepted that the near-infrared (NIR) and X-ray flares in Sagittarius A*, the compact radio source associated 
with a mass $M\sim 3-4\times 10^6\;M_\odot$ supermassive black hole at the Galactic Center (Sch\"{o}del et al. 2002; Ghez et 
al. 2004; the Schwarzschild radius $r_S =10^{12}(M/3.4\times 10^6\;M_\odot$) cm), are produced by relativistic 
electrons within a few Schwarzschild radii of the black hole via synchrotron and synchrotron self-Comptonization (SSC) 
processes, respectively (Baganoff et al. 2001, 2003; Genzel et al. 2003; Belanger et al. 2006; Eckart et al.  2004, 2006; 
Yusef-Zadeh et al. 2006a). 
Studying the details of the electron acceleration therefore plays an important role in revealing the properties of the 
flaring plasma and its interaction with the surroundings and the space-time created by the black hole, which may lead to a 
measurement of the spin with comprehensive flare observations and general relativistic MHD and ray-tracing 
simulations of the black hole accretion and the corresponding radiation transfer.

The electrons are likely accelerated by turbulent plasma waves generated in an accretion torus via instabilities, 
which induce the release of gravitational energy of the plasma (Balbus \& Hawley 1991; Tagger \& Melia 2006). In the simplest 
case, in which there is no large scale magnetic field, the turbulent magnetic field $B$ can play the dual role of both 
heating the electrons
(presumably via resonant wave particle interactions) and cooling them (via SSC radiation), since the heating and cooling rates have different energy dependencies. Earlier, we showed that the {\it steady-state} electron spectrum can be 
approximated as a relativistic Maxwellian distribution with the ``temperature'' $\gamma_cm_ec^2$ depending on the plasma 
density $n$ and the coherent length of the magnetic field, which should be comparable to the size of the flare region $R$. 
Simultaneous spectroscopic observations in the NIR and X-ray bands (with four measured quantities: fluxes and spectral 
indexes in both bands) can then be used to determine the four basic model parameters, namely $B$, $R$, $\gamma_c m_ec^2$ and 
$n$, and to test the model (Liu et al. 2006a, 2006b).

These investigations, however, also indicated that the steady-state solutions are not always applicable, especially for NIR 
flares with very soft spectra. A time-dependent approach to the evolution of the electron distribution under the influence 
of the turbulent magnetic field is required to study the flares in quantitative details. In light of the recent 
discovery of a correlation between the flux $F_\nu\propto\nu^{\alpha-1}$ and the spectral index $\alpha$ in the NIR band 
(Ghez et al. 2005; Gillessen et al. 2006; Krabbe et al. 2006; Hornstein et al. 2006), we carry out such an investigation.
We show that relativistic Maxwellian distributions in general give a good description of the electron spectra with the 
corresponding ``temperature'' evolution determined by the initial temperature, the magnetic field and its coherent length and 
the gas density. It is therefore possible to couple these results with MHD simulations to study the radiative characteristics 
of collisionless plasmas self-consistently.

Several scenarios have been suggested to explain the observed correlation between flux and spectral index (Gillessen et al. 
2006). We find here that the dynamical processes of plasma heating and cooling are the most likely mechanism and the 
observation requires that the magnetic field be anti-correlated with the electron temperature. Since the evolution of the 
magnetic field depends on the turbulence generation mechanism, which needs to be addressed via MHD simulations, we calculate 
the evolution of the electron distribution function and the corresponding radiation spectrum for a cooling and a heating 
phase, where the magnetic field is set to be inversely proportional to the mean electron energy. The model naturally recovers 
the observed correlation between the NIR flux and spectral index and predicts the X-ray emission characteristics accompanying 
the NIR flares. Simultaneous flare observations in the NIR and X-ray bands can readily test the model and uncover the 
underlying physical processes producing these flares.

The kinetic equation of the electron acceleration and its time-dependent solutions are described in \S\ \ref{acc}. In \S\ 
\ref{apl} the model is applied to flares in Sagittarius A* by calculating the radiation from a flaring plasma with 
prescribed magnetic field evolutions. The implication and limitation of these results are discussed in \S\ \ref{dis}.

\section{Time Dependent Solutions of Electron Acceleration by Plasma Waves}
\label{acc}

The theory of electron acceleration by a turbulent magnetic field in a hot collisionless plasma has been discussed 
by Liu et al. (2006b). Regardless the details of resonant wave-particle coupling, it was shown that the evolution of the 
electron distribution function $N(\gamma,t)$ can be described by the following kinetic equation
\begin{equation} 
   {\partial N \over \partial t}
=  {\partial \over \partial \gamma}\left[{\gamma^2\over \tau_{\rm ac}}{\partial N \over \partial 
\gamma} + \left({\gamma^2\over \tau_0}-{2\gamma\over \tau_{\rm ac}}\right)N\right]\,,
\label{kinetic}
\end{equation}
where the synchrotron cooling and acceleration times are given, respectively, by
\begin{eqnarray}
\tau_{\rm syn}(\gamma) &=&\tau_0/\gamma \equiv {9m_e^3c^5/4 e^4B^2\gamma}\,, \\
\tau_{\rm ac} &\equiv& 2\gamma^2/<\Delta\gamma\Delta\gamma/\Delta t> = C_1{3R c/v_{\rm A}^{2}}
= {12\pi C_1 n m_p cR/ B^2}\,,
\label{tac}
\end{eqnarray} 
the Alfv\'{e}n velocity $v_{\rm A} = B/(4\pi n m_p)^{1/2}$, $C_1$ is a dimensionless quantity 
of order 1, and we have assumed that there is no particle escape and therefore no return current in association with the 
electron acceleration \footnote{These processes are not expected to affect the results significantly as far as the escape time 
scale is much longer than $\tau_{\rm ac}$ (Liu et al. 2006a).}. The size of the flaring region, the magnetic field and gas 
density are given respectively by $R$, $B$ and $n$. The scattering mean free path of the electrons, $C_1 R$, is comparable to 
the source size. $m_e$, $m_p$, $c$, $e$, and $\gamma$ are the electron mass, proton mass, speed of light, elemental charge 
unit and the Lorentz factor of the electron, respectively.

To demonstrate the behavior of the time-dependent solutions, we consider the simplest case, where $\tau_{\rm ac}$ and 
$\tau_0$ remain constant in time, and normalize the distribution function. Then the steady state is $N(\gamma)= (\gamma^2/ 
2\gamma_c^3)\exp(-\gamma/\gamma_c)\,, \, {\rm with} \ \gamma_c = \tau_0/\tau_{\rm ac}=3m_e^3c^4/16\pi C_1 e^4m_p\ n\ R$. Here 
the integral over $\gamma$ has been extended from 1 to 0 to simplify the expression. We note that $\gamma_c$ is independent 
of $B$ because both the synchrotron cooling and acceleration rates are proportional to $B^2$, while $R$ and $n$ control 
$\gamma_c$ by affecting the turbulence heat rate via modification of the coherent length and Alfv\'{e}n velocity $v_{\rm A}$, 
respectively. Once one specifies the initial electron distribution with $\tau_{\rm ac}$ as the time unit, the solution only 
depends on $\tau_0$ or the steady-state temperature $\gamma_c$.

\begin{figure}[bht] 
\begin{center}
\includegraphics[height=8.4cm]{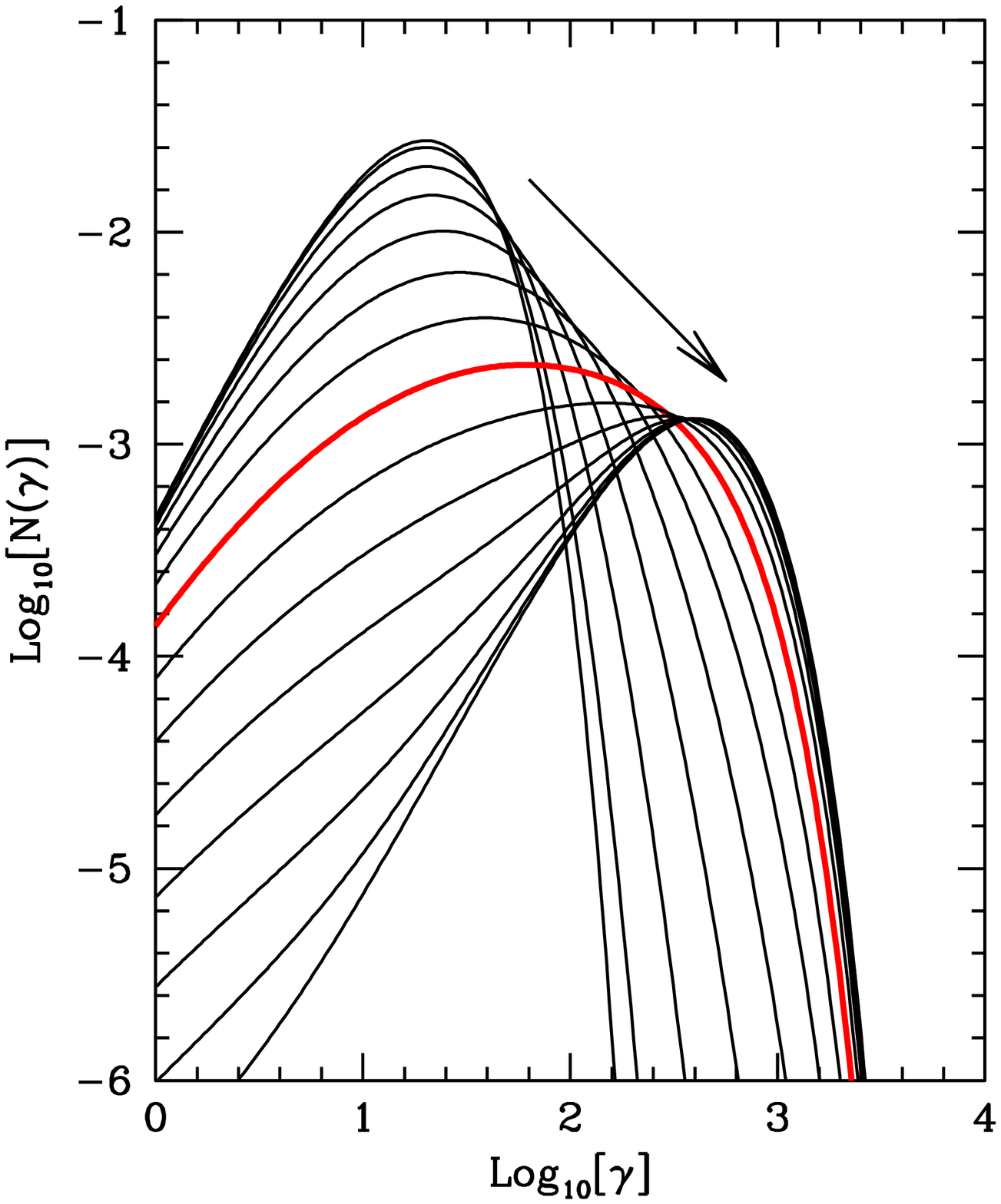}
\hspace{-0.6cm}
\includegraphics[height=8.4cm]{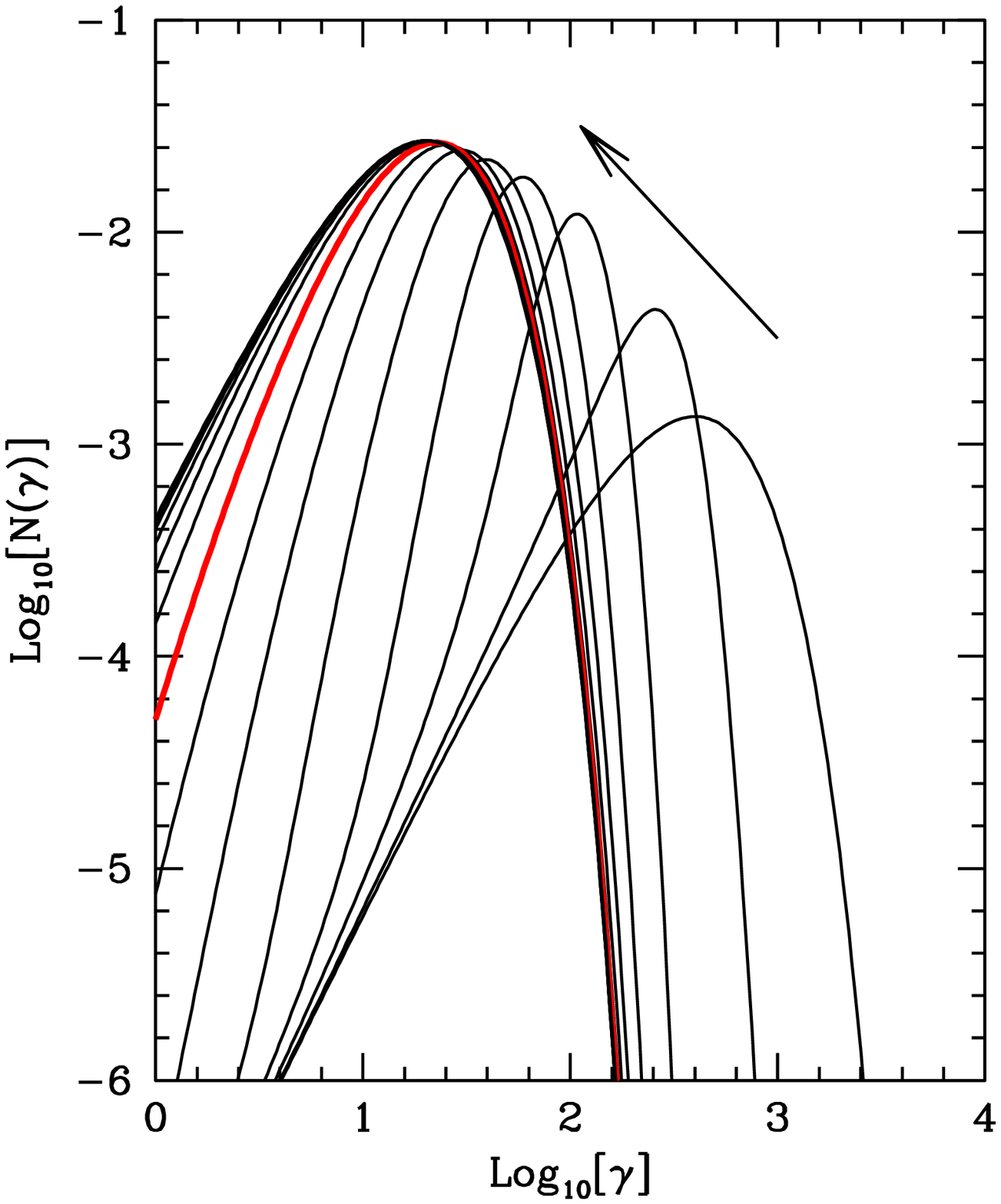}
\end{center}
\caption{
{\it Left:} Evolution of the electron distribution function $N(\gamma)$. The initial and final temperatures (in unit of 
$m_ec^2$) are $10$ and $200$, respectively. The time steps are $4(i/14)^2\tau_{\rm ac}$, where $i$ is an integer ranging from 
0 to 14. The arrow indicates the direction of the evolution. The thick line is for $i=7$ corresponding to the time step 
$\tau_{\rm ac}$. We note that the electron distribution is already very similar to the steady-state spectrum after $\tau_{\rm 
ac}$.
{\it Right:} Same as the left panel except the initial and final temperatures exchanged. Note the pileup of electrons during 
the first few time steps due to the SSC cooling.}
\label{fig1.ps}
\end{figure}

In what follows, the energy unit is assumed to be $m_ec^2$. Figure \ref{fig1.ps} shows the evolution of $N(\gamma, t)$ 
starting with a relativistic Maxwellian distribution with an initial temperature $\gamma_i$ until $t=4\tau_{\rm ac}$. The 
left panel has $\gamma_i=10$ and $\gamma_c = 200$, which corresponds to a heating phase of the plasma. We note that the 
particle distributions are already close to the steady-state solution after the time step $t=\tau_{\rm ac}$. The right panel 
has $\gamma_i=200$ and $\gamma_c=10$ corresponding to a cooling phase. Because the synchrotron cooling (with a timescale 
$\gamma_c\tau_{\rm ac}/\gamma$) dominates above $\gamma_c$, the particle distribution evolves faster than in the heating 
phase (in unit of $\tau_{\rm ac}$). We also note the pileup of electrons to a nearly monotonic distribution in the early 
section 
of this cooling phase, due 
to more rapid cooling of higher energy electrons. Later the distribution is broadened by the diffusion term in the kinetic 
equation (the first term on the right hand side of eq. [\ref{kinetic}]).

\begin{figure}[bht] 
\begin{center}
\includegraphics[height=8.4cm]{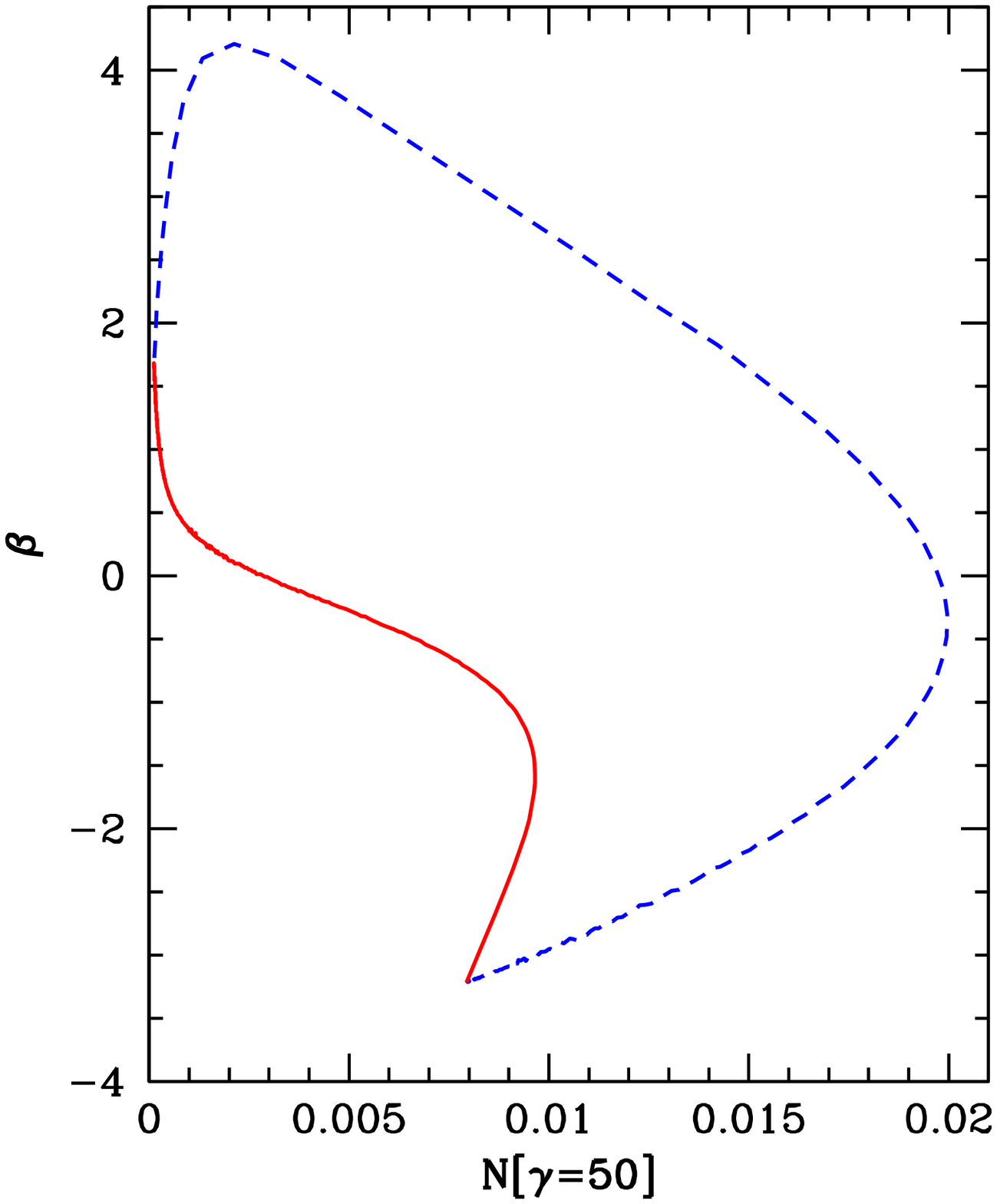}
\hspace{-0.6cm}
\includegraphics[height=8.4cm]{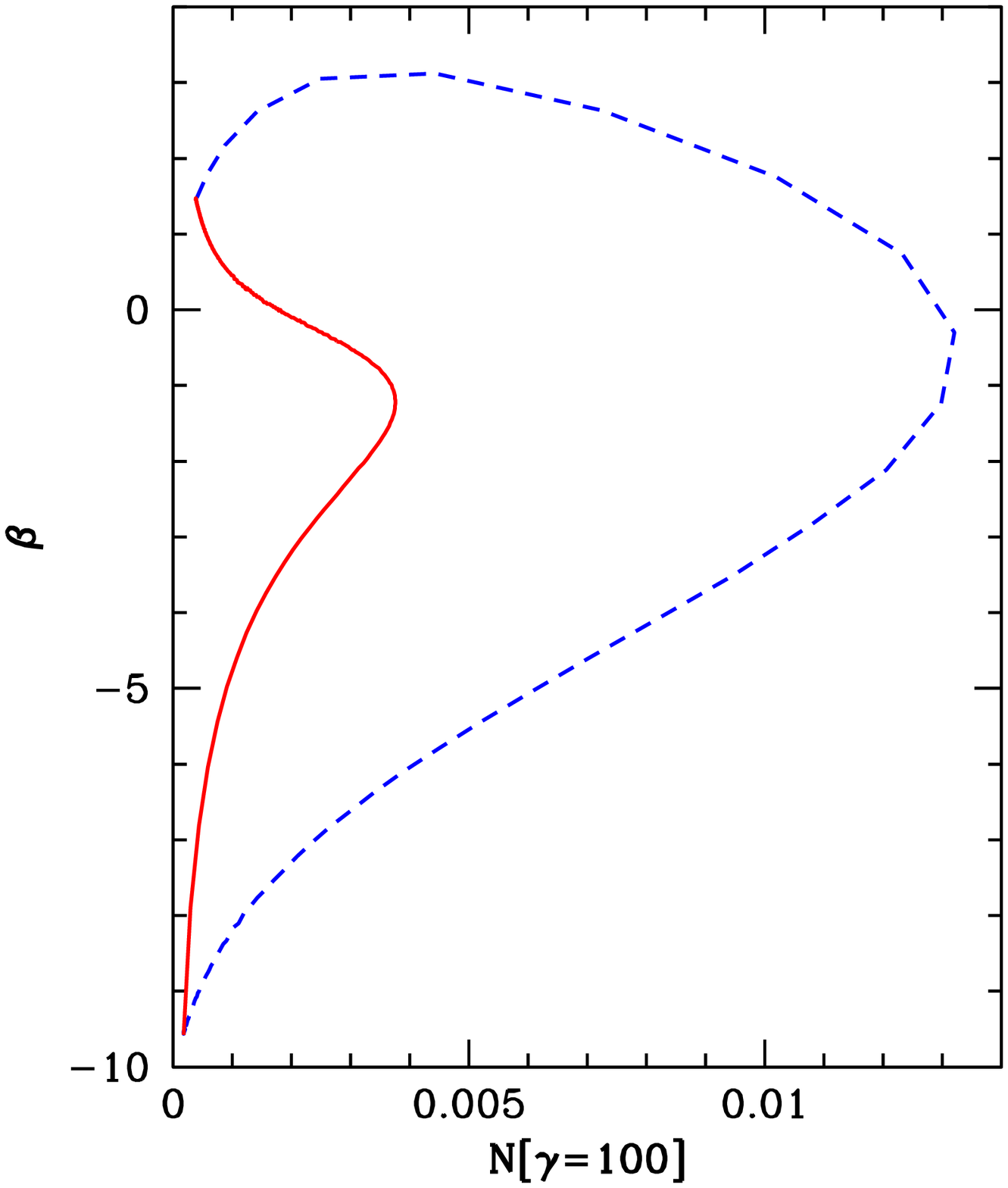}
\end{center}
\caption{
{\it Left:} Correlation between $N$ and $\beta=\d \ln{N}/\d\ln{\gamma}$ at $\gamma=50$. The solid line is for the heating 
phase. The dashed line is for the cooling phase. The rising of the spectral index to above 2 is caused by the pileup of 
electrons in the early cooling phase. The correlation is quite different for the two phases.
{\it Right:} Same as the left panel but for $\gamma=100$.
}
\label{fig2.ps}
\end{figure}

Figure \ref{fig2.ps} shows the correlations between $N(\gamma, t)$ and the electron spectral index $\beta\equiv \partial 
\ln{N(\gamma, t)}/\partial \ln{\gamma}$ at $\gamma=50$ (left) and $100$ (right) for the two runs in Figure \ref{fig1.ps}. 
Here the solid and dashed lines are for the heating and cooling phases, respectively (one may consider the time $t$ as a 
parametric variable). The rising of $\beta$ to above 2 is due to the pileup of electrons in the early cooling phase. The 
correlations are quite different for the two phases and energies. For a given magnetic field, these correlations mimic 
similar correlations in the synchrotron radiation, which we investigate in \S\ \ref{apl}.


To characterize the evolution of $N(\gamma, t)$ in general, we next consider the evolution of the mean energy $<\gamma>(t) = 
\int \gamma N(\gamma, t) \d \gamma/\int N(\gamma, t)\d \gamma$. Because the heating of electrons by turbulence depends on the 
derivative of $N(\gamma, t)$ with respect to $\gamma$ via the diffusion term, the evolution of $<\gamma>$ can be very complicated. Fortunately for any smooth function $N(\gamma)$ 
one may approximate this derivative term with a heating term, i.e. the total energy change rate 
\begin{equation}
\dot{<\gamma>} \simeq -{<\gamma>^2\over \tau_{\rm ac}\gamma_c}+{(2+a)<\gamma>\over \tau_{\rm ac}}\,,
\label{heq}
\end{equation}
where a ``\ $\dot{}$\ '' indicates a derivative with respect to time and $a$ is a parameter to be determined by the steady 
state solution. Then we have
\begin{equation}
<\gamma>_t={(2+a)\gamma_c\gamma_i\over \gamma_i-(\gamma_i-\gamma_c)\exp[-(2+a)t/\tau_{\rm ac}]}\,.
\label{mgamt}
\end{equation}
In the steady state, $<\gamma>_t=3\gamma_c$, therefore $a=1$.

\begin{figure}[bht] 
\begin{center}
\includegraphics[height=8.4cm]{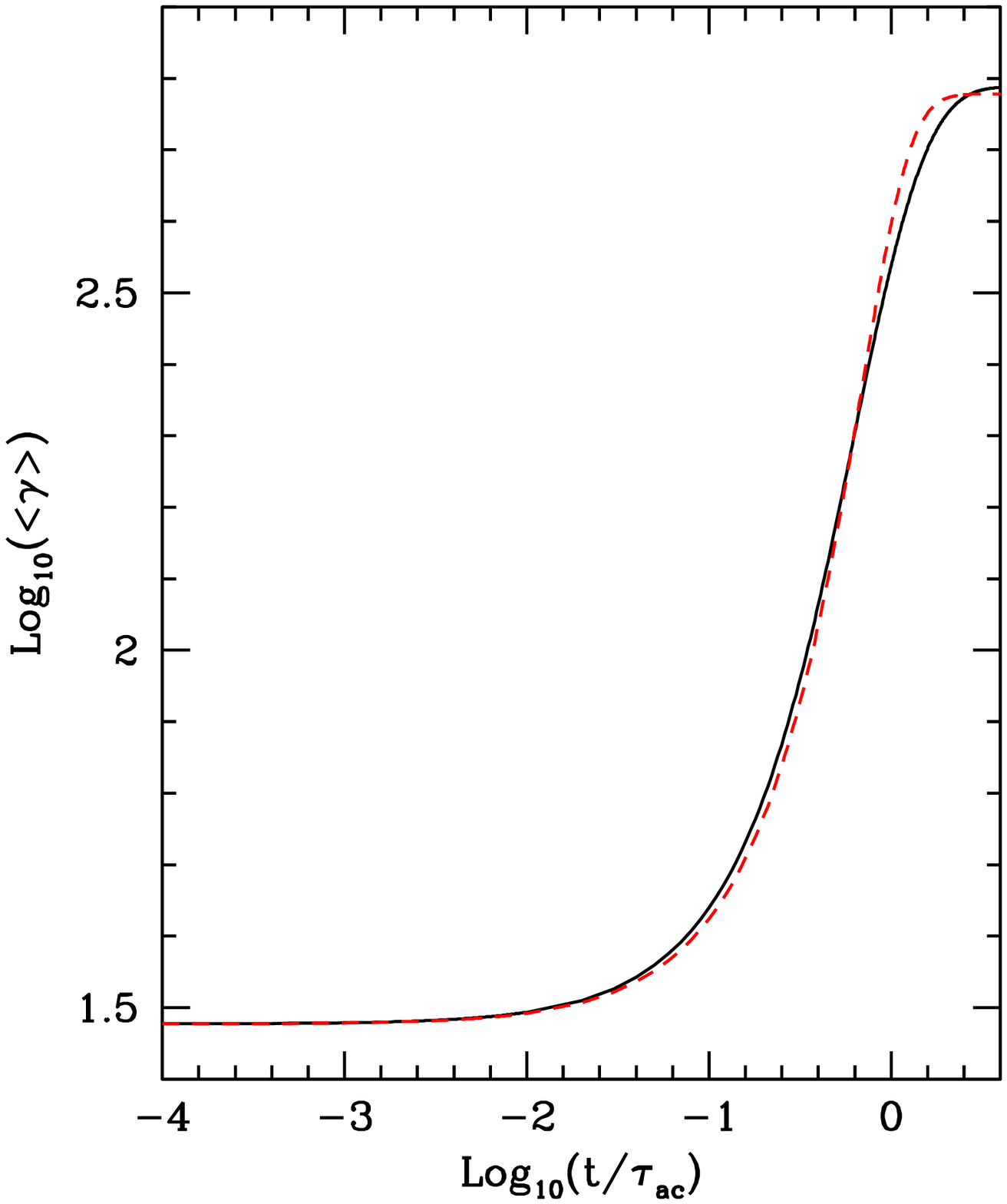}
\hspace{-0.6cm}
\includegraphics[height=8.4cm]{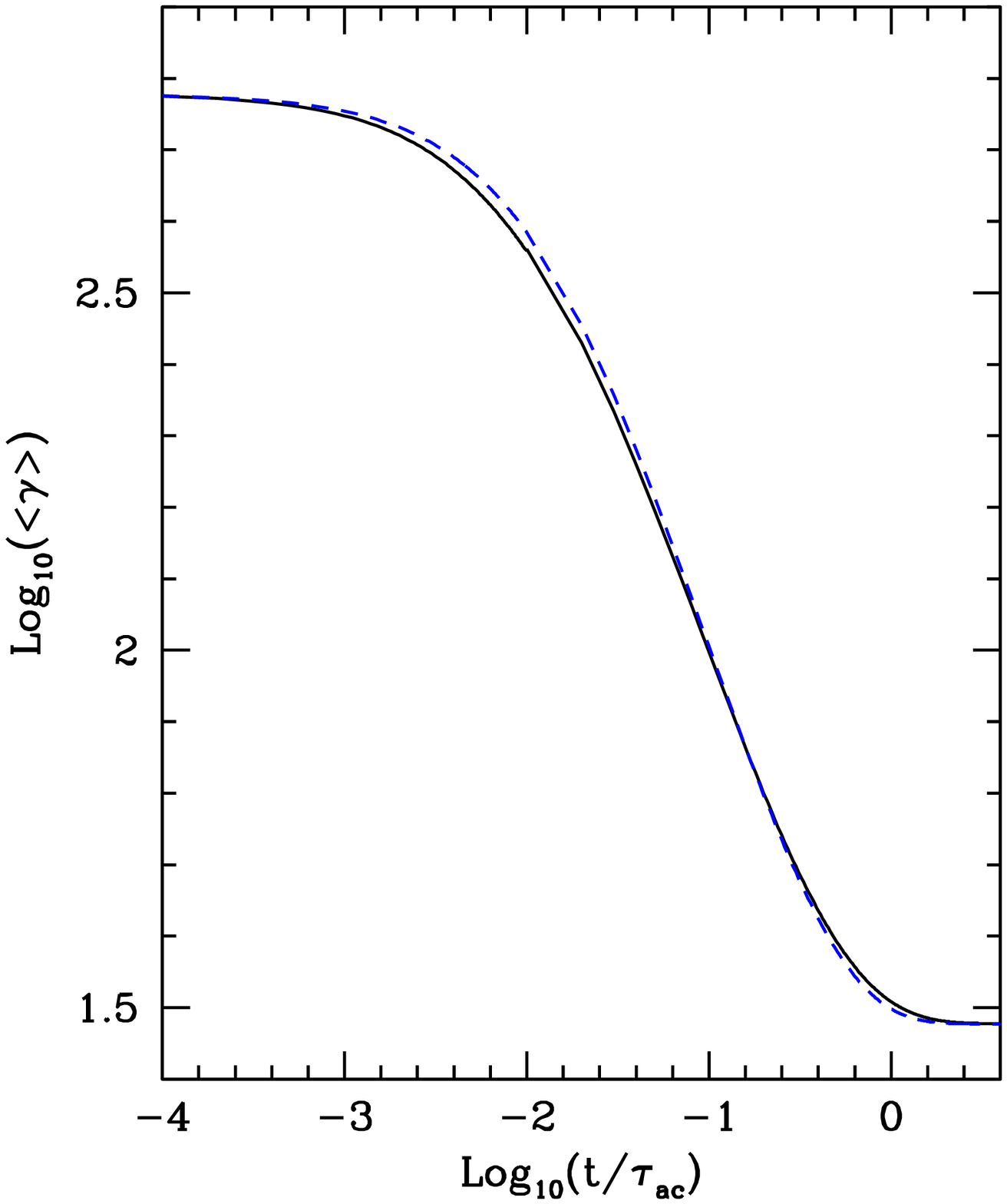}
\end{center}
\caption{
{\it Left:} Evolution of the mean energy of the electrons in the heating phase. The dashed line indicates the theoretical 
result given by equation (\ref{mgamtf}) with $a=1.6$.
{\it Right:} Same as the left panel but for the cooling phase with $a=1.0$.
}
\label{fig3.ps}
\end{figure}

\begin{figure}[bht] 
\begin{center}
\includegraphics[height=8.4cm]{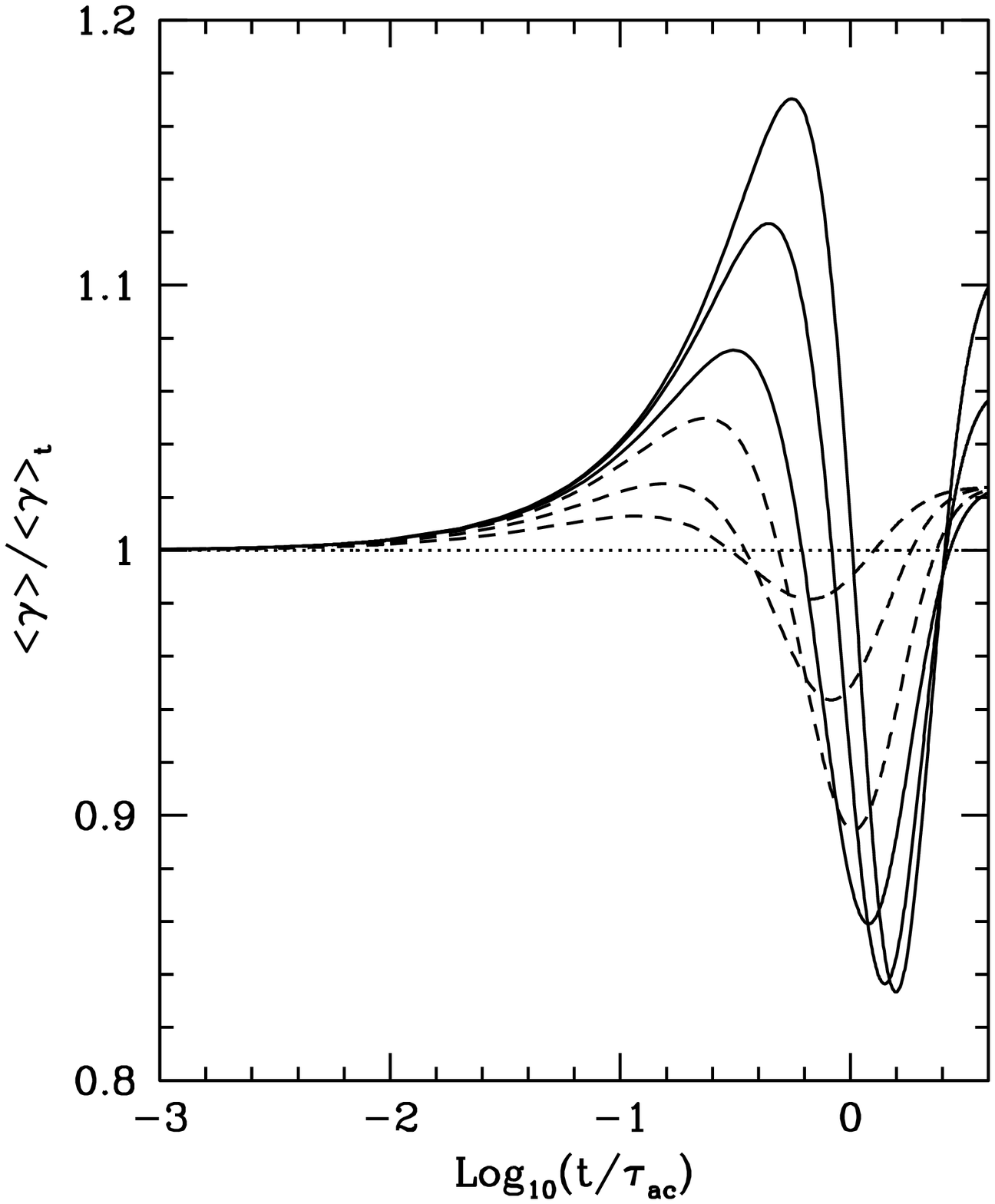}
\hspace{-0.6cm}
\includegraphics[height=8.4cm]{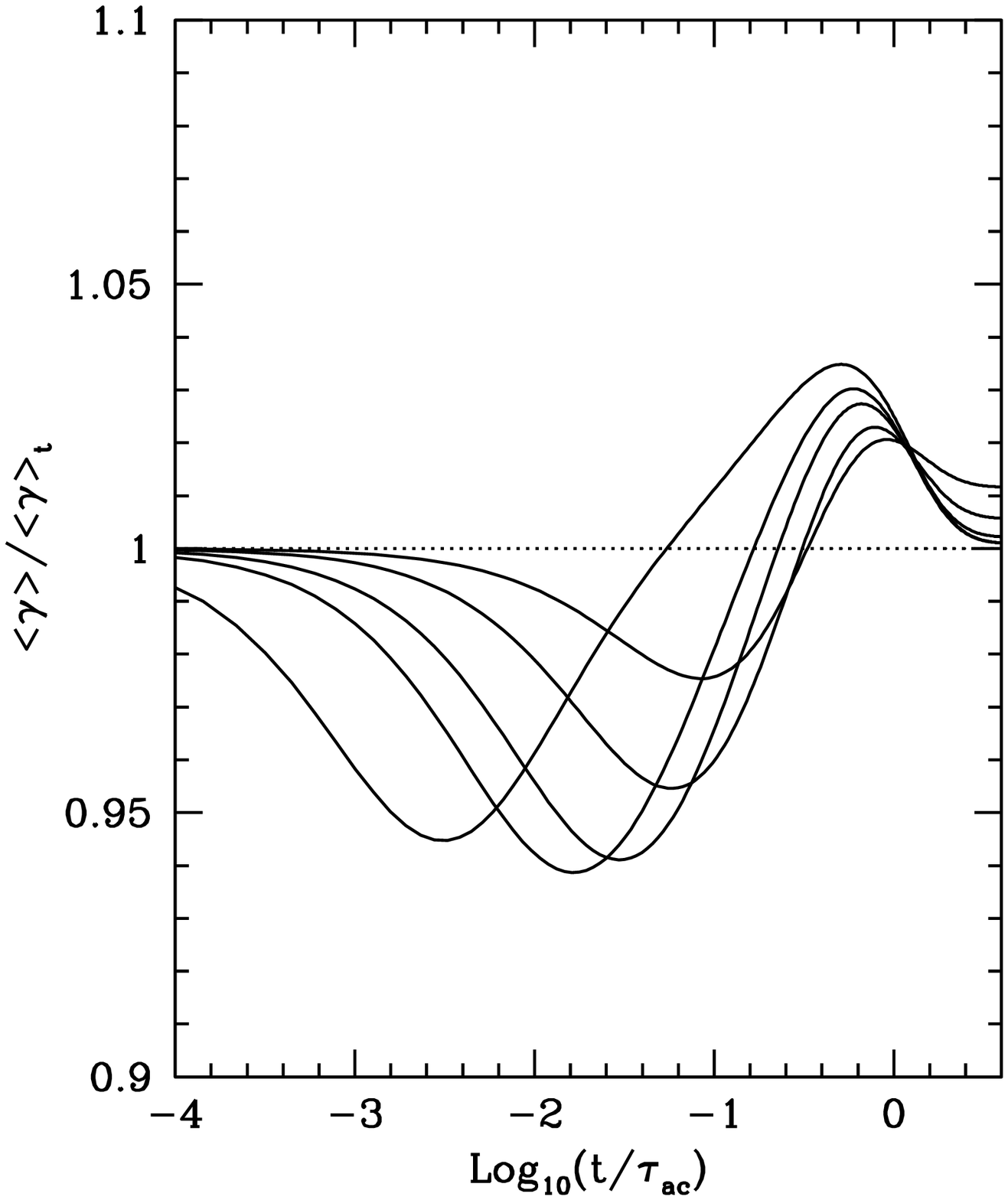}
\end{center}
\caption{
{\it Left:} The ratio of the mean energy of the electrons to the theoretical value as a function of time for several initial  
and final temperatures in the heating phase. The solid lines have an initial temperature of $10$ and final temperatures of 
200, 500, 1000. The dashed lines have a final temperature of 200 and initial temperatures of 20 50, and 100. The relative 
error (the difference between the numerical and theoretical results divided by the theoretical values) increases with the 
increase of the ratio of the final to initial temperature (the dynamical range). 
{\it Right:} Same as the left panel but for the cooling phase. The line with the greatest relative error corresponds to a 
cooling from  1000 to 10. The other lines have an initial temperature of 200 and final temperatures of 10, 20, 50, and 100. 
The relative error also increases with the dynamical range. 
}
\label{fig4.ps}
\end{figure}

However the above approximation is not good enough for the intermediate 
steps of the heating phase when the particles have a broad distribution 
(Fig. \ref{fig1.ps}) and the heating by the diffusion term is more 
efficient. To fit the numerical results we set the ``$a$'' in the 
numerator of equation (\ref{mgamt}) equal to one to be consistent with the 
steady state solution and leave the ``$a$'' in the denominator as a free 
parameter:
\begin{equation}
<\gamma>_t={3\gamma_c\gamma_i\over \gamma_i-(\gamma_i-\gamma_c)\exp[-(2+a)t/\tau_{\rm ac}]}\,.
\label{mgamtf}
\end{equation}
We find that $a=1.6$ and $1.0$ give good fits to the $<\gamma>$ evolutions
of the heating and cooling phases, respectively.
Figure \ref{fig3.ps} shows the evolution of $<\gamma>$ and the corresponding fits for the heating 
(left) and cooling (right) phases. Figure \ref{fig4.ps} gives the ratio of $<\gamma>$ to the fit values
for several initial and final temperatures. The relative error is within $20\%$ for the heating phase 
and within $10\%$ for the cooling. The error also increases with the increase of the dynamical range, 
which is the ratio of the initial temperature to steady state temperature for the cooling phase and vice versa
for the heating phase. 
Equation (\ref{mgamtf}) therefore gives a good description of the evolution of $<\gamma>$ under the 
influence of a turbulent magnetic field. 

We note that equation (\ref{mgamtf}) can be generalized to address the heating and cooling of collisionless plasma by 
turbulent plasma waves. The heating time $\tau_{\rm ac}$ (see eq.[\ref{tac}]) usually is a function of time.  Taking into 
account the effects discussed in the previous paragraph, equation (\ref{mgamt}) leads to 
\begin{equation}
<\gamma>_t={3\gamma_c\gamma_i\over \gamma_i-(\gamma_i-\gamma_c)\exp[-(2+a)\int\d t/\tau_{\rm ac}(t)]}\,.
\label{mgamtfg}
\end{equation}
MHD simulations can give the time evolution of $B$, its coherent length, and $n$. Equation (\ref{mgamtfg}) can then be 
used in these simulations to address the heating and cooling of electrons. In \S\ \ref{apl} we study the 
SSC emission of these electrons during flares in Sagittarius A*.

\section{Synchrotron Emission and SSC}
\label{apl}

We consider the case where the particle distribution is isotropic with respect to the turbulent magnetic field. 
Then the synchrotron flux density and emission coefficient at frequency $\nu$ are 
given, respectively, by (Pacholczyk 1970)
\begin{equation}
F_{\nu}(\nu)={4\pi R^3\over 3 D^2}{\cal E}_\nu\,,\ \ \ \
{\cal E}_\nu(\nu)={\sqrt{3} e^3\over 4\pi m_e c^2} B\,n\, \int_0^\infty\d \gamma
\int_0^1 \d \mu (1-\mu^2)^{1/2} N(\gamma) F(x) 
\,,
\end{equation}
where 
\begin{eqnarray}
x&=&{\nu\over \nu_c}\equiv {4\pi m_e c\ \nu\over 3eB(1-\mu^2)^{1/2}\gamma^2}\,,
\\ 
F(x) &=& x\int_x^\infty K_{5/3}(z)\d z \simeq {2.1495 x^{1/3}e^{-2x}}+1.348 x^{1/2}e^{-x-0.249x^{-1.63}}\,,
\end{eqnarray}
$D$, $\mu$, and $K_{5/3}$ are the distance to the Galactic Center, the cosine of the angle between the magnetic field and line 
of sight, and the corresponding Bessel function, respectively, and the approximation for $F(x)$ is accurate within $8\%$. We 
therefore have the spectral index in a given narrow frequency range $\alpha \equiv {{\rm d}\ln(\nu F_\nu)/ {\rm d}\ln\nu}$. 

As shown by Liu et al. (2006b), most of the synchrotron radiation is emitted in the optically thin region for flares in 
Sagittarius A*. For a uniform spherically symmetric source, the self-Comptonization flux density is then given by (Blumenthal 
\& Gould 1970)
\begin{eqnarray}
F_X(\nu) \simeq {\pi e^4\nu n R\over 6 m_e^2 c^4}\int_0^\infty 
\d\gamma
{N(\gamma)\over \gamma^4}
\int_{\nu\over 4\gamma^2}^\infty 
\d \nu^\prime
{F_\nu(\nu^\prime)\over \nu^{\prime3}}
\left(2\nu\ln{\nu\over 4\gamma^2\nu^\prime}+\nu+4\gamma^2\nu^\prime-{\nu^2\over 2\gamma^2\nu^\prime}\right)
\,,
\end{eqnarray}
Similarly one can define the X-ray spectral index $\alpha_X \equiv {{\rm d}\ln(\nu F_X)/ {\rm d}\ln\nu}\,.$

\subsection{Flux and Spectral Index Correlations with a Constant Magnetic Field}
\label{conb}

\begin{figure}[bht] 
\begin{center}
\includegraphics[height=8.4cm]{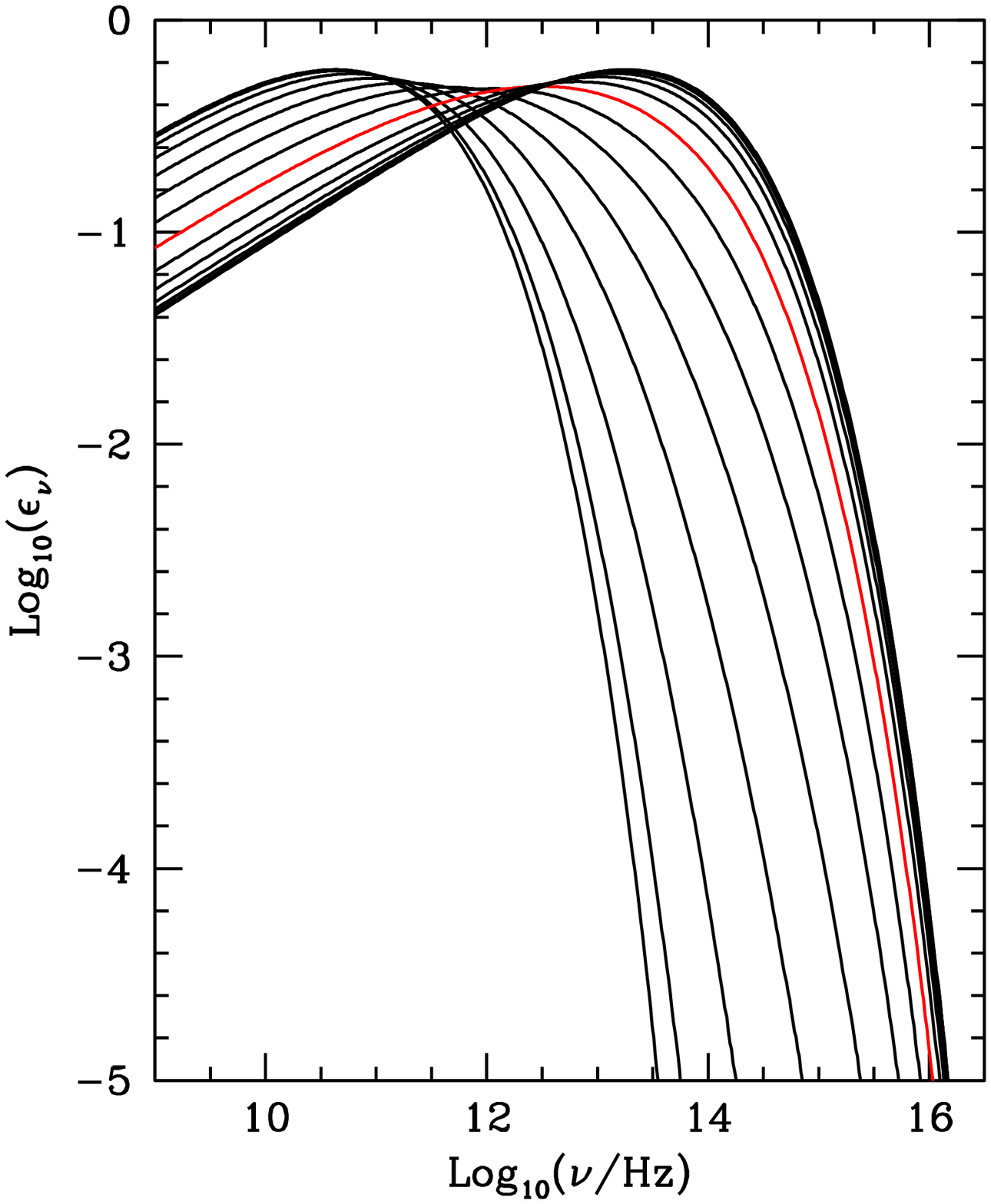}
\hspace{-0.6cm}
\includegraphics[height=8.4cm]{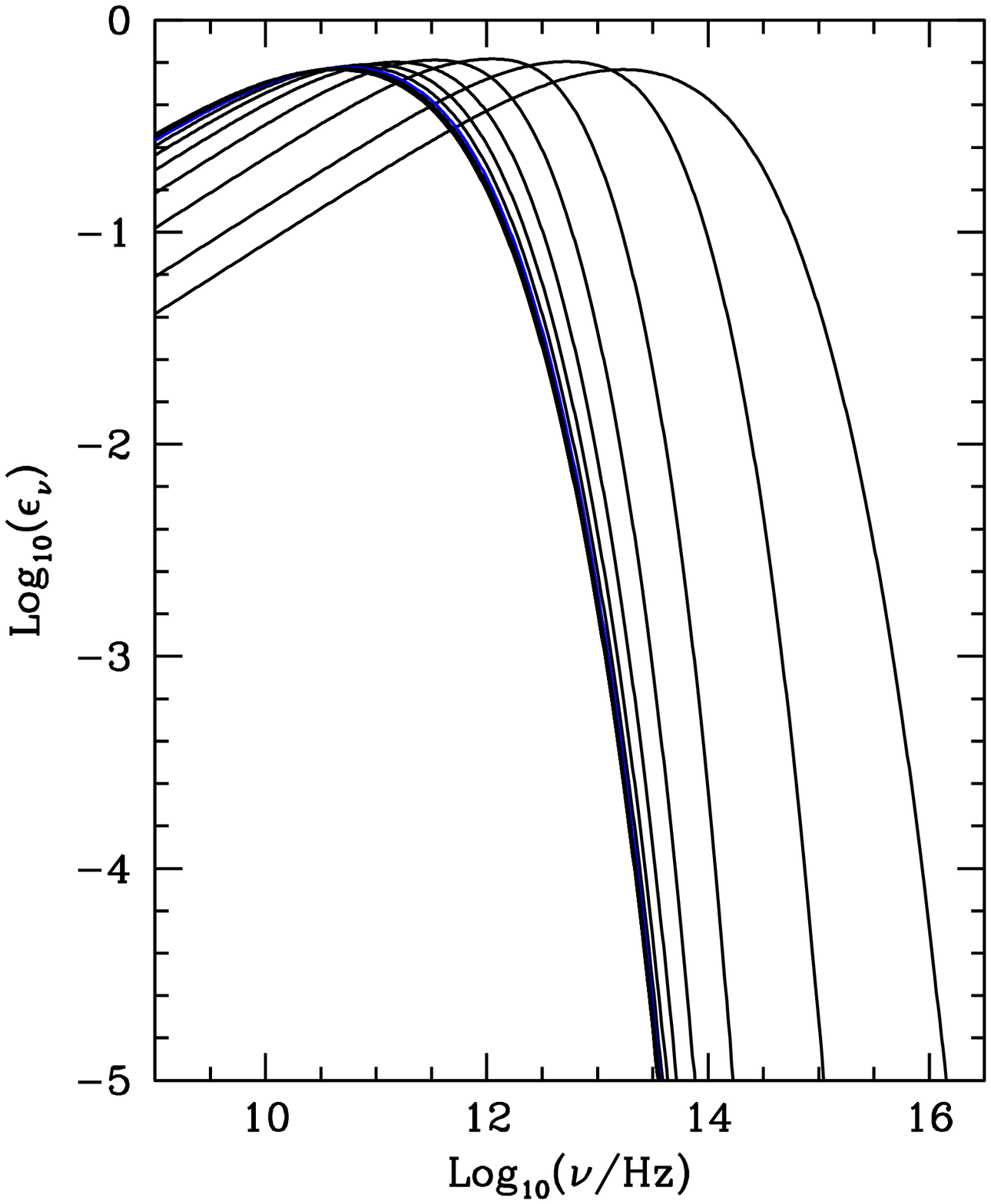}
\end{center}
\caption{
Evolution of the normalized synchrotron flux density spectrum $\epsilon_\nu$ for the two runs shown in Figure 
\ref{fig1.ps}. The time steps are the same as in Figure \ref{fig1.ps} and the magnetic field $B=100$ G. The left and 
right 
panels correspond to the heating and cooling phases, respectively. The evolution of $\epsilon_\nu$ is less dramatic than that 
of $N(\gamma)$ due to the dominance of synchrotron emission by more energetic electrons. The spectra are very similar to 
thermal synchrotron spectra.
}
\label{fig5.ps}
\end{figure}

To demonstrate how the heating and cooling processes affect the evolution of the radiation spectrum, we first consider the 
cases where the magnetic field $B$ and the total number of electrons $4\pi nR^3/3$ remain constant. Figure \ref{fig5.ps} 
shows the evolution of the normalized synchrotron flux density spectrum
\begin{equation}
\epsilon_\nu(\nu, t) \equiv {F_\nu(\nu, t) \sqrt{3} m_e c^2 D^2\over e^3 B n R^3}=\int_0^\infty\d \gamma
\int_0^1 \d \mu (1-\mu^2)^{1/2} N(\gamma, t) F(x)
\label{epss}
\end{equation}  
for the two runs in Figure \ref{fig1.ps} with $B=100$ G. Compared to the evolution of $N(\gamma)$, the evolution of 
$\epsilon_\nu(\nu)$ is less dramatic because the synchrotron spectrum is dominated by emission from the more energetic 
electrons, and the radiation spectra are very similar to thermal synchrotron spectra. However, the difference between the 
heating and cooling phases is obvious.

\begin{figure}[bht] 
\begin{center}
\includegraphics[height=8.4cm]{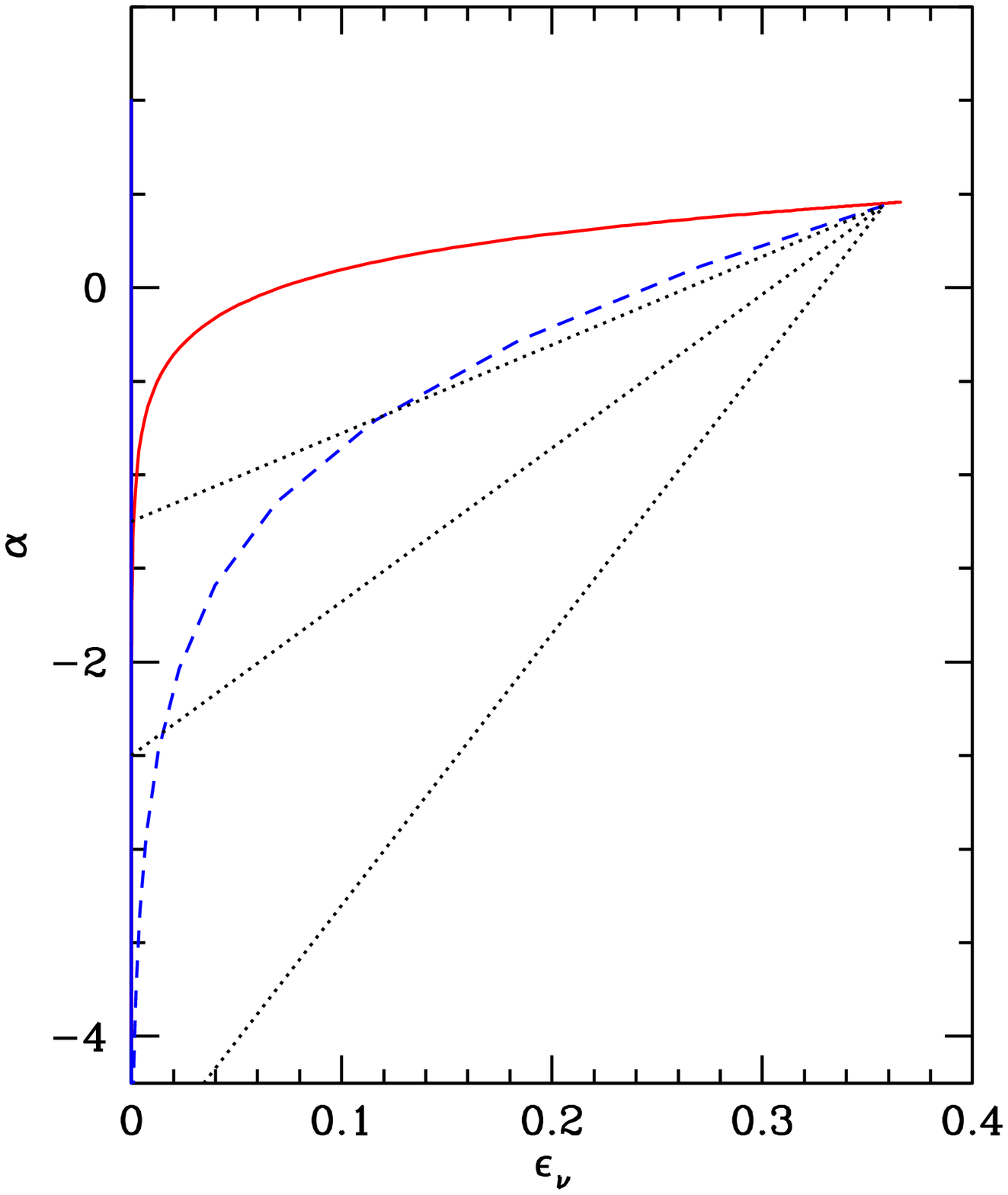}
\hspace{-0.6cm}
\includegraphics[height=8.4cm]{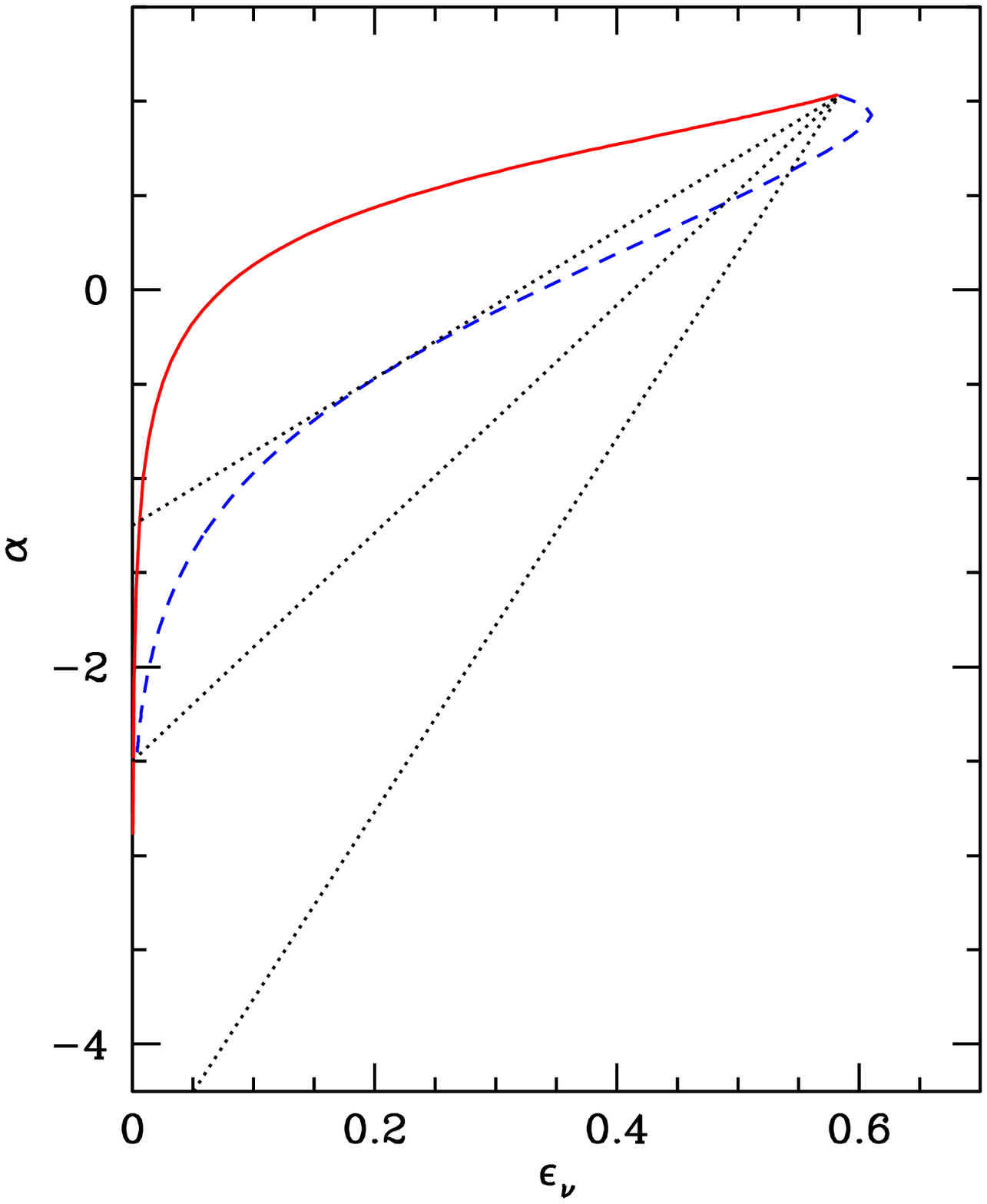}
\end{center}
\caption{Correlation between the normalized flux density $\epsilon_\nu$ and the spectral index $\alpha$ at 
$\nu=1.4\times10^{14}$ Hz (left) and $1.4\times10^{13}$ Hz (right) for the two runs in Figure \ref{fig5.ps}. 
The solid and dashed lines are for the heating and cooling phases, respectively. The dotted lines give the observational 
results. The cooling phase correlation fits the observations marginally, while the heating phase correlation is too flat 
compared with observations. See text for details.
}
\label{fig6.ps}
\end{figure}

Figure \ref{fig6.ps} shows the correlations between $\alpha$ and $\epsilon_\nu$ at $\nu=1.4\times 10^{14}$ Hz (left) and 
$1.4\times 10^{13}$ Hz (right). The dotted lines indicate the observed correlations with different background 
subtraction methods (Gillessen et al. 2006). The cooling phase fits the observations marginally, while the heating phase 
predicts a correlation much weaker than is observed. We note that for constant $B$ and $nR^3$ the observed flux 
density is proportional to $\epsilon_\nu$. The correlation between $\epsilon_\nu$ and $\alpha$ can be compared with 
observations directly by adjusting the normalization factor $BnR^3$. We also note that for given $N(\gamma, t)$ the 
correlation between $\epsilon_\nu$ and $\alpha$ only depends on $\nu/B$. Therefore for the same runs the correlations at 
$1.4\times10^{14}$ Hz with $B=1000$ G will be the same as those shown in the right panel.

In \S\ \ref{acc} we showed that the evolution of $N(\gamma, t)$ is very similar in the heating and cooling phases for 
different initial and final temperatures, and that adjusting these temperatures changes the starting and ending points of the 
correlation in the $\epsilon_\nu$-$\alpha$ plane. The shape of the correlation, however, does not change dramatically. We 
therefore conclude that for a constant (and uniform) magnetic field the observed correlation between flux and spectral index 
in the NIR band can marginally be attributed to the cooling of a plasma heated up instantaneously at the onset of the 
flare.

Adiabatic expansion or compression and Doppler boosting have also been suggested to explain the correlations between spectral 
index and flux during flares (Yusef-Zadeh et al. 2006b; Gillessen et al. 2006). To produce NIR flares with very soft spectra 
the electrons producing the NIR emission need to have a sharp high energy cutoff. As shown by Liu et al. (2006a, 2006b) and 
discussed in \S\ \ref{acc}, a relativistic Maxwellian distribution gives the most natural approximation to the electron 
spectrum. For a temperature of $\gamma_0$ (Liu et al. 2006b) we have, 
\begin{equation}
\epsilon_\nu(\nu) = 0.5 x_M I(x_M)\,,
\end{equation}
where
\begin{eqnarray}
I(x_M)&=& 4.0505x_M^{-1/6}(1+0.40x_M^{-1/4} + 0.5316x_M^{-1/2})\exp(-1.8899\,x_M^{1/3})\,,
\label{Im}
\\
x_M &=& {\nu/\nu_0}
\equiv {4\pi m_e c\nu/3 e B \gamma_0^2} 
\,,
\label{xm}
\end{eqnarray}
and
\begin{equation}
\alpha = 1.833 - 0.6300
x_M^{1/3}-{0.1000x_M^{1/4}+0.2658\over
x_M^{1/2}+0.4000x_M^{1/4}+0.5316}\,.
\label{alpha}
\end{equation}

The left panel of Figure \ref{fig7.ps} shows the correlation between $\epsilon_\nu$ and $\alpha$. As mentioned above, when 
the magnetic field is constant, this can be compared with observations directly. Such a correlation can be produced when a 
plasma is expanding or compressed adiabatically in a uniform magnetic field so that the temperature $\gamma_0$ changes. 
The theoretical prediction is clearly inconsistent with the observed results. A very unusual electron spectrum is needed to 
make the model of adiabatic expansion in a uniform magnetic field in line with observations.

Because ${\cal E}_\nu/\nu^2$ is a Lorentz invariant, $F_\nu/\nu^2$ is approximately a constant under the Lorentz transform. 
The correlation between $F_\nu$ and $\alpha$ due to Doppler boosting effects is therefore given by the correlation between 
$\epsilon_\nu/x_M^2$ and $\alpha$, which is much flatter than the correlation between $\epsilon_\nu$ and $\alpha$  and 
therefore can not explain the observations.

We conclude that in a constant and uniform magnetic field only SSC cooling of a hot plasma heated instantaneously is 
marginally in agreement with the observed spectral index and flux correlation in the NIR band. Turbulent heating, adiabatic 
processes and Doppler effects produce a correlation which is much flatter than the observed results. 

\subsection{Flux and Spectral Index Correlations with a Variable Magnetic Field}
\label{vmag}

\begin{figure}[bht] 
\begin{center}
\includegraphics[height=8.4cm]{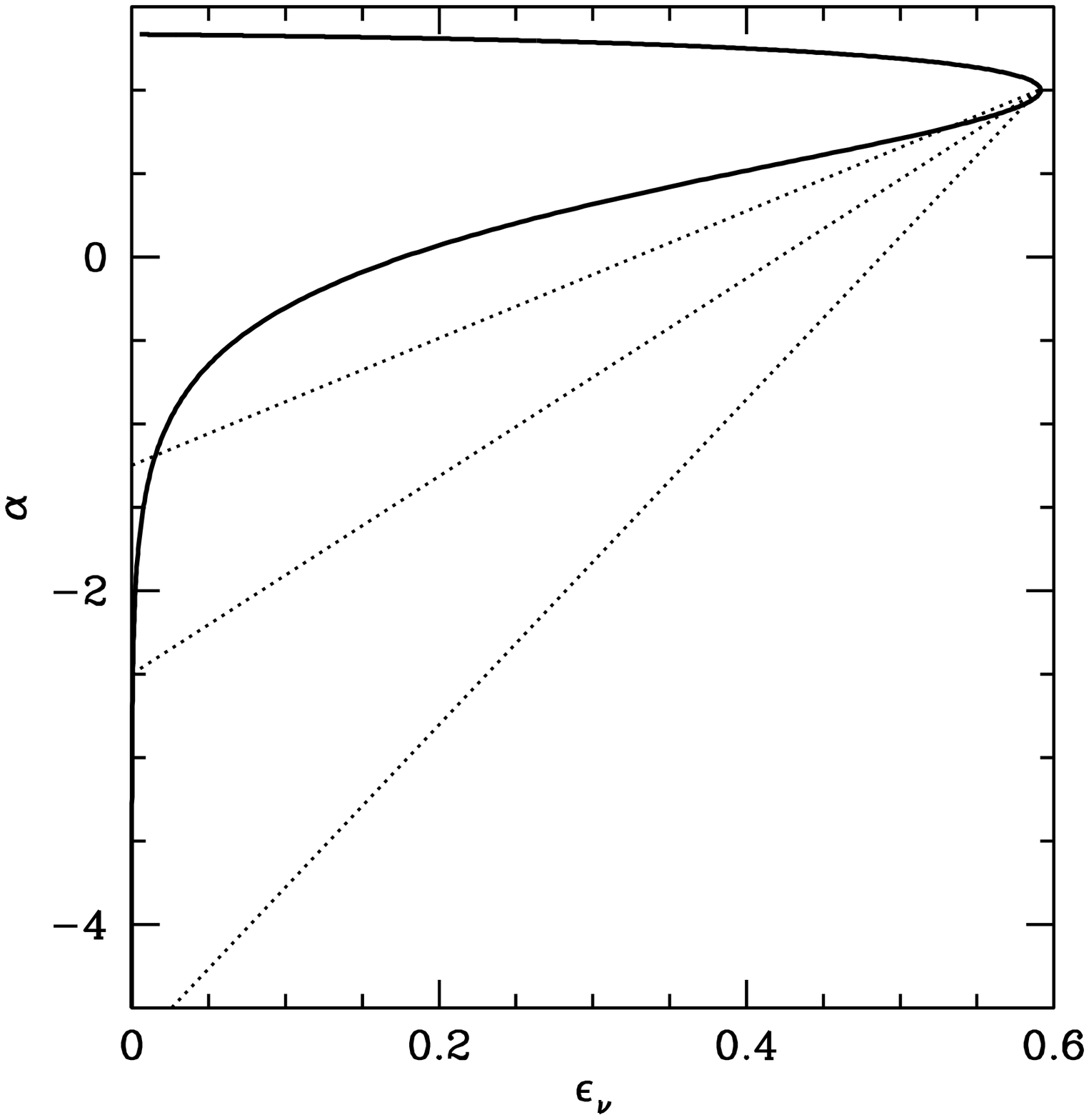}
\hspace{-0.6cm}
\includegraphics[height=8.4cm]{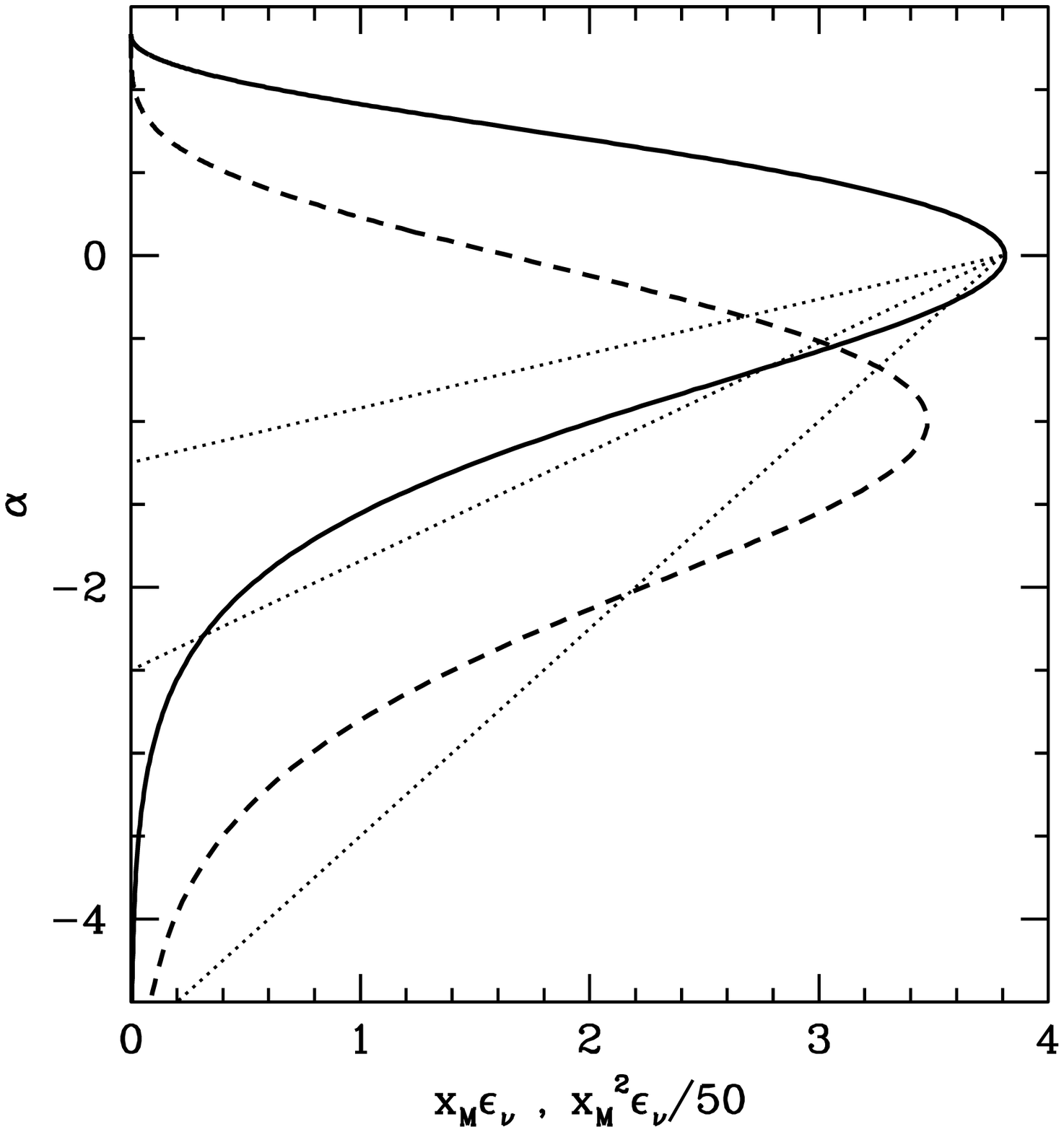}
\end{center}
\caption{
{\it Left:} Correlation between $\epsilon_\nu$ and $\alpha$ for electrons with a relativistic Maxwellian distribution. The 
dotted lines indicate the observed results, which clearly lie below the theoretical curve.
{\it Right:} Correlations between $x_M\epsilon_\nu$ (solid line), $x_M^2\epsilon_\nu/50$ (dashed line) and $\alpha$, which 
are more consistent with observations. See text for details.
}
\label{fig7.ps}
\end{figure}

To improve the model fitting to the observed correlations, we next consider the scenario where the magnetic field is 
variable. The correlation between flux and spectral index is depicted by the correlation between $B\epsilon_\nu$ and 
$\alpha$. For a relativistic Maxwellian distribution with a temperature $\gamma_0$, if $B\gamma_0$ is a constant in time, 
$B\epsilon_\nu$ is proportional to $x_M\epsilon_\nu$. As shown in the right panel of Figure \ref{fig7.ps}, the correlation of 
the solid line gives a better fit to the observations than all models discussed above. We note, however, that for $\alpha$ 
increasing from -4 to 1, the temperature of the plasma has to increase by about three orders of magnitude (Liu et al. 2006b).

In the more general cases where $B\gamma_0^p$ (with $p$ as a constant) does not change with time, $B\epsilon_\nu$ is 
proportional to $x_M^{p/(2-p)}\epsilon_\nu$. For $p=0$, we recover the result with a constant magnetic field. It is clear 
that $p$ has to be positive and less than 2 to make the flux and spectral index correlation steeper, implying an 
anti-correlation between the magnetic field and the mean electron energy during the flare evolution. The dashed line in 
Figure \ref{fig7.ps} (right) corresponds to $p=4/3$, which fits the correlation in the soft- and dim- state but predicts a 
spectrum, which is softer than the observed ones when the flux is high. On the other hand, the recent re-analyses of the Keck 
observations of flares from Sagittarius A* appear to favor such a scenario (Krabbe et al. 2006).

To take into account the heating and cooling effects, we choose a magnetic field inversely proportional to $<\gamma>$ during 
the evolution of the flare, which makes the heating and cooling rates depend on the electron distribution. Compared with the 
heating and cooling processes studied in \S\ \ref{acc}, one more parameter is needed to obtain the solutions: the initial magnetic field. This parameter also sets the characteristic timescale of the system, which is related to the 
initial value of $\tau_{\rm ac}$. Using an explicit number scheme with the time steps scaled as $\tau_{\rm ac}(t)$, we 
calculate the evolution of the synchrotron spectrum for a cooling and heating phase.

\begin{figure}[bht] 
\begin{center}
\includegraphics[height=8.4cm]{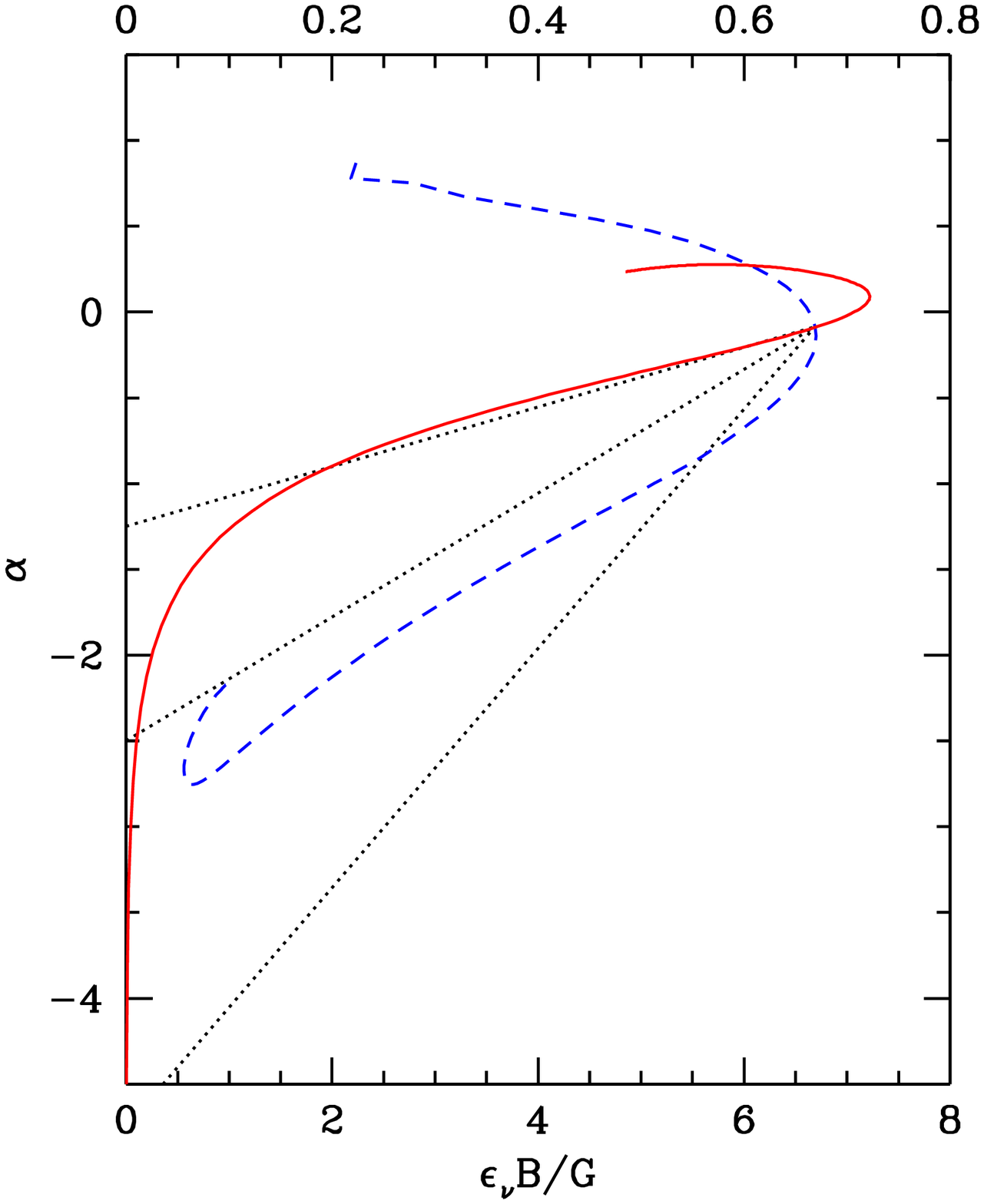}
\hspace{-0.6cm}
\includegraphics[height=8.4cm]{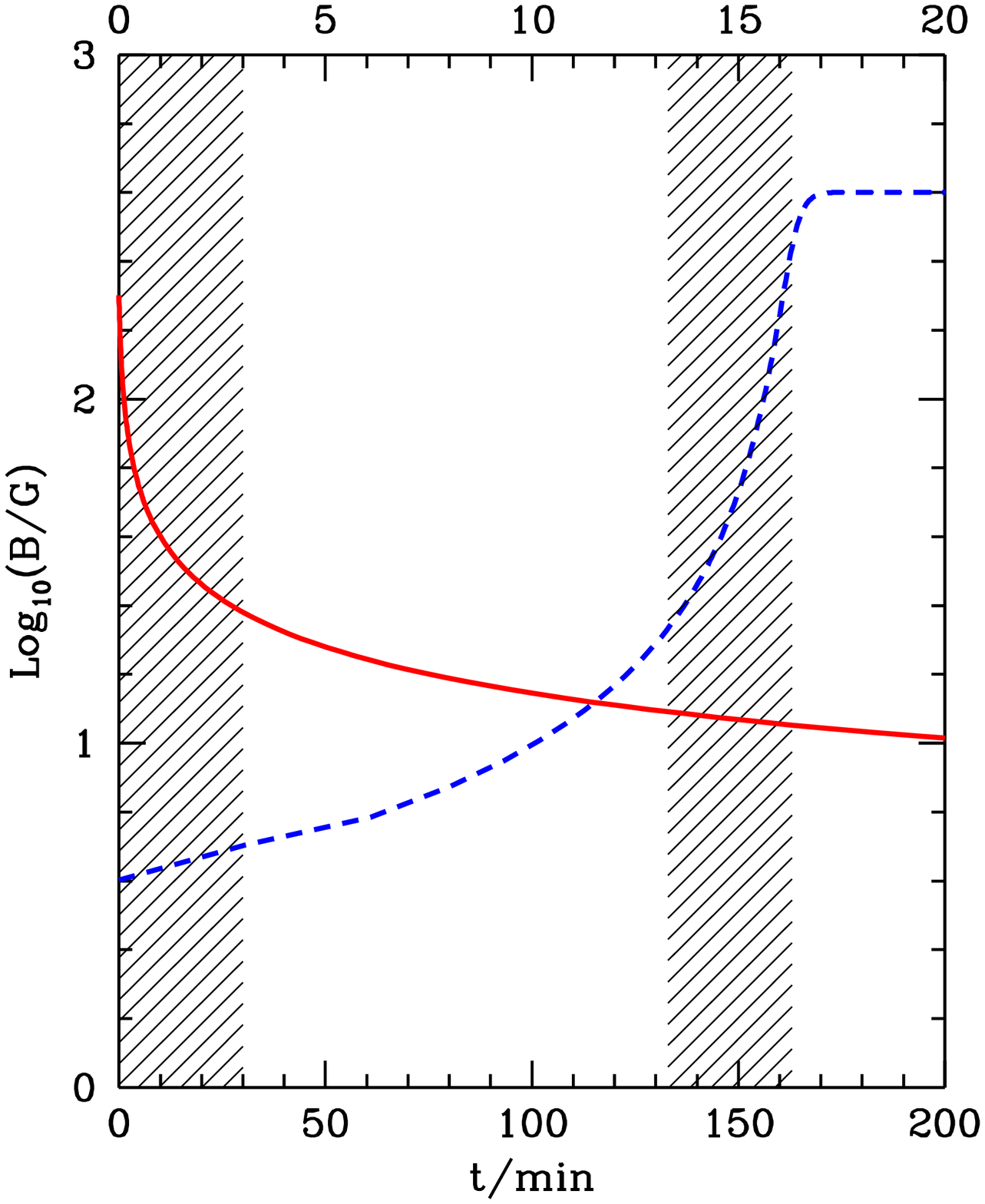}
\end{center}
\caption{
{\it Left:} 
Correlation between the flux $\epsilon_\nu B$ and spectral index $\alpha$ at $1.4\times 10^{14}$ Hz for the heating (solid 
line) and cooling (dashed line) phases, where the magnetic field $B$ is chosen to be inversely proportional to the mean energy 
of the electrons $<\gamma>$. The initial magnetic field, the initial and final temperatures are $4$ Gauss, 2000 and 20 and 
200 Gauss, 10 and 1000 for the cooling and heating phases, respectively. The dotted lines indicate the observational results. 
The scale of the heating phase flux is indicated on the upper axis.
{\it Right:} 
Time evolution of the magnetic field for the heating (solid line; upper scale) and cooling (dashed line; lower scale) phases.
The regions most relevant to flare (0 to 3 minutes for the heating phase and 133 to 163 minutes for the cooling phase) 
observations are shaded.
}
\label{fig8.ps}
\end{figure}

The dashed line in Figure \ref{fig8.ps} (left) shows the correlation between $\epsilon_\nu B$ and $\alpha$ for the cooling 
phase at $\nu = 1.4\times 10^{14}$ Hz, which is consistent with the observations. The lower magnetic field, the initial and 
final temperatures, are $4$ Gauss, 2000 and 20, respectively. Then we have $n R =2.1\times10^{19} C_1^{-1}$ cm$^{-2}$ and 
$\tau_{\rm ac} = 6.5 \times 10^{3} (B/10\ {\rm G})^{-2}$ mins. Here a relatively large temperature change is chosen to produce 
significant variation of the synchrotron spectral index. The dashed line in the right panel gives the evolution of the 
magnetic field. Although the initial magnetic field is relatively low, the steady state magnetic field is $400$ Gauss, which 
is much higher than the typical value for the quiescent state. 

\begin{figure}[bht] 
\begin{center}
\includegraphics[height=8.4cm]{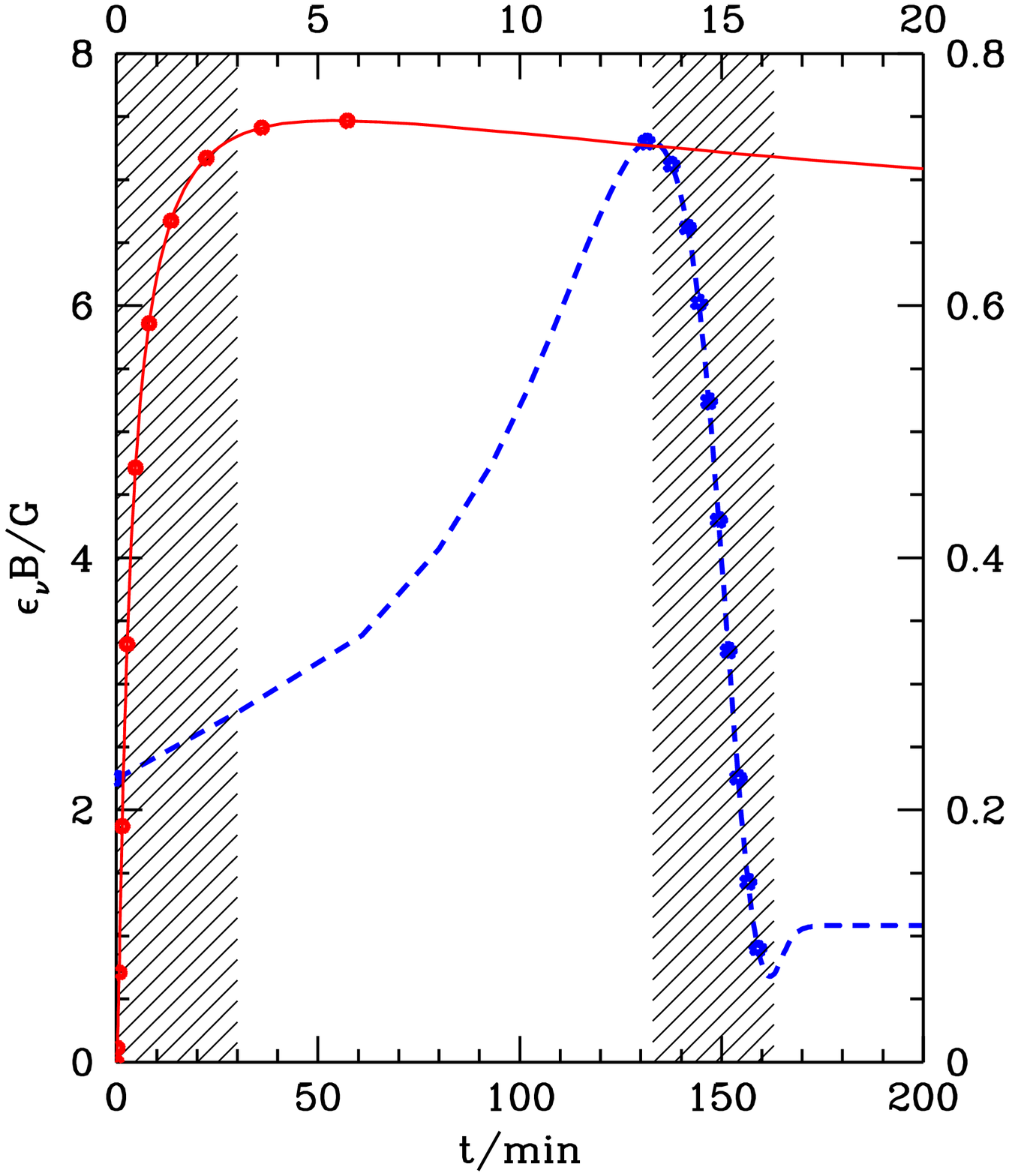}
\hspace{-0.6cm}
\includegraphics[height=8.4cm]{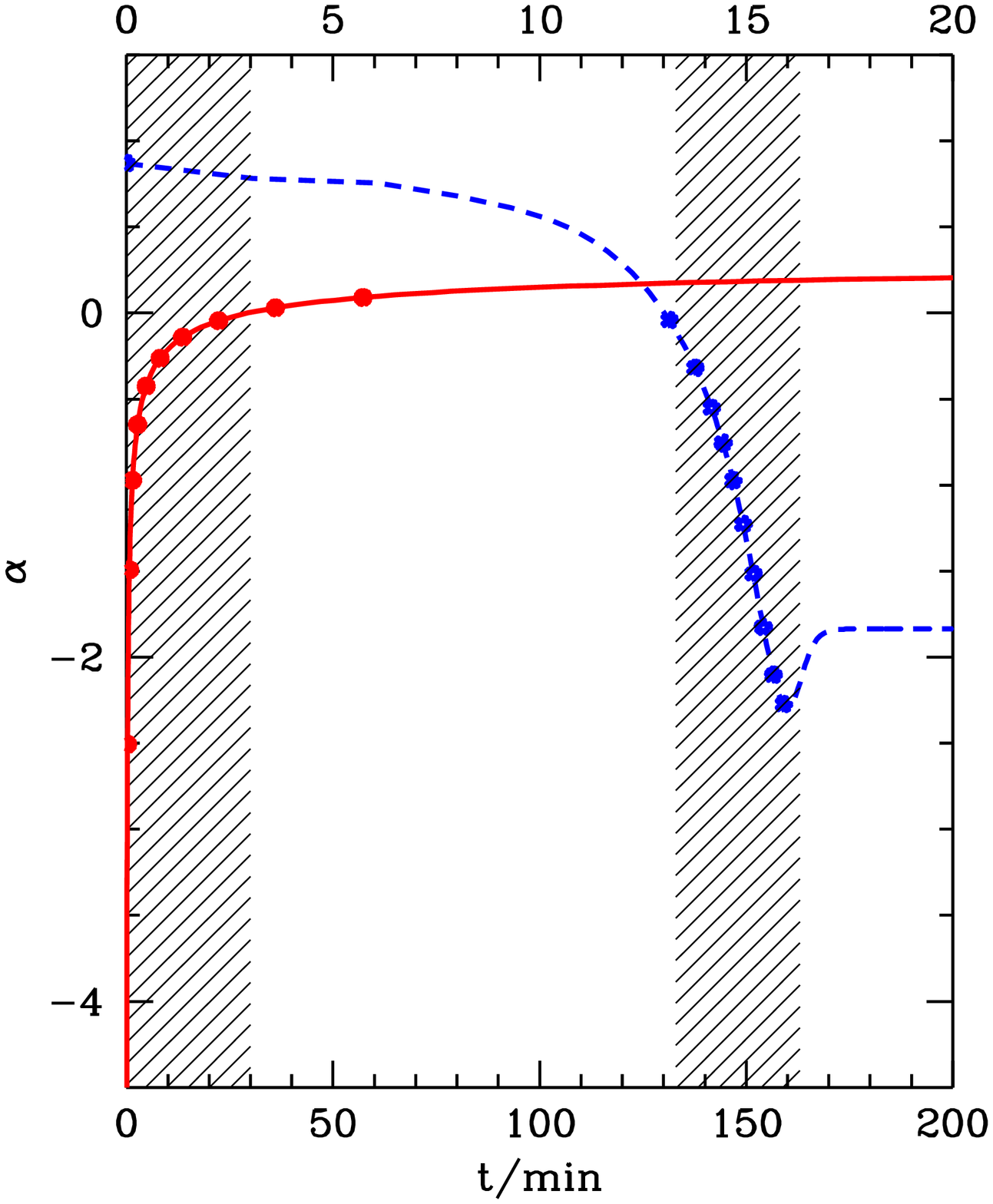}
\end{center}
\caption{
Same as the right panel of Figure \ref{fig8.ps} but for the flux (left) and spectral index (right) at $1.4\times 10^{14}$ Hz. 
The scale of the heating 
phase flux is indicated on the right axis.  Solid circles correspond to the NIR spectra in Figure \ref{fig10.ps}. See text for details.}
\label{fig9.ps}
\end{figure}

The dashed lines in Figure \ref{fig9.ps} show the light curves of the flux (left) and spectral index (right) for the cooling phase. We note that 
the correlation between the flux and spectral index mostly occurs between 130 to 160 minutes. Before 130 minutes, the system 
evolves slowly due to the relatively lower magnetic field ($< 20$ Gauss). Because the magnetic field in the accretion torus 
in Sagittarius A* is about a few tens of Gauss in the quiescent state, the evolution before 100 minutes is likely irrelevant 
to flares in Sagittarius A* since the magnetic field is below 10 Gauss during this period. From 100 to 170 minutes the model 
predicted correlation is quite in line with observations. The model not only produces the correct correlation slope, but also 
recovers the typical variation timescale of a few tens of minutes. After 170 minutes, the system reaches a steady state. 
The dips near $160$ mins in the flux and spectral index light curves are related to the turn around of the spectral index 
and flux correlation before reaching the steady-state. These are caused by the sharp cutoff (sharper than an exponential 
cutoff) of the electron spectra in the 
cooling phase, which makes the spectra softer than that of a relativistic Maxwellian distribution. 
To produce a flare state with $\alpha = -0.5$ and $F_\nu = 5$ mJy, the total number of electrons 
involved in the flare $4\pi n R^3/3 = 2.6 \times 10^{41}$, giving rise to $R = 5.5 \times 10^{10} C_1^{1/2}$ cm and $n=3.7 
\times 10^{8} C_1^{-3/2}$ cm$^{-3}$. These values agree with results of previous studies (Liu et al. 2006a, 2006b).

The heating phase fits the observed correlation marginally (the solid line in the left panel of Figure \ref{fig8.ps}). The 
initial magnetic field, the initial and final temperatures are $200$ Gauss, 10 and 1000, respectively, giving rise to $n R 
=4.1\times10^{17} C_1^{-1}$ cm$^{-2}$ and $\tau_{\rm ac} = 1.3 \times 10^{2} (B/10\ {\rm G})^{-2}$ mins. The solid lines in 
the right panel of Figure \ref{fig8.ps} and in Figure \ref{fig9.ps} depict the evolution of the magnetic field and the flux 
and spectral index, respectively. The heating phase proceeds much faster than the cooling phase due to the relatively high 
initial magnetic field and final temperature. The flux reaches the peak value within $\sim 6$ minutes, after which the 
spectral index saturates near 0.2, and both the magnetic field and flux decrease slowly. To produce even harder spectra, one 
has to increase the value of $B \gamma_c^2$ at the steady state dramatically (Liu et al. 2006b), making the heating phase 
proceed on an even shorter timescale ($\propto 1/B^2\gamma_c$). A very weak initial magnetic field ($<1$ Gauss) and a 
very high temperature ($\gg1000$) are required to produce hard NIR spectra and a flux rising time of a few minutes.
To produce a flare state with $\alpha = -0.5$ and $F_\nu = 5$ mJy, the total number of electrons involved in the flare $4\pi 
n R^3/3 = 4.7 \times 10^{42}$, which is more than ten times larger than that in the cooling phase.  This leads to $R = 1.6 
\times 10^{12} C_1^{1/2}$ cm and $n=2.5 \times 10^{5} C_1^{-3/2}$ cm$^{-3}$. 

In general the spectral index $\alpha$ of the above models is always less than $4/3$. To produce even harder spectra, one 
has to introduce self-absorption effects. For an electron temperature $\gamma_0\simeq 200$, the size of the emission 
region is then given by $$R\simeq (F_\nu /2\pi \gamma_0 m_e)^{1/2} D/\nu = 3.7\times 10^7 (D/8 {\rm kpc})(F_\nu/5\ {\rm 
mJy})^{1/2}(\gamma_0/200)^{-1/2} (\nu/1.4\times 10^{14}{\rm Hz})^{-1} {\rm cm}\,,$$ which is much smaller than any relevant 
length scales. We therefore do not expect NIR flares with spectral indexes larger than $4/3$. Any solid detection of very 
hard NIR spectra will rule out this model and may be in conflict with any synchrotron models. 

\subsection{Flux and Spectral Index Correlation in the X-ray Band}

\begin{figure}[bht] 
\begin{center}
\includegraphics[height=8.4cm]{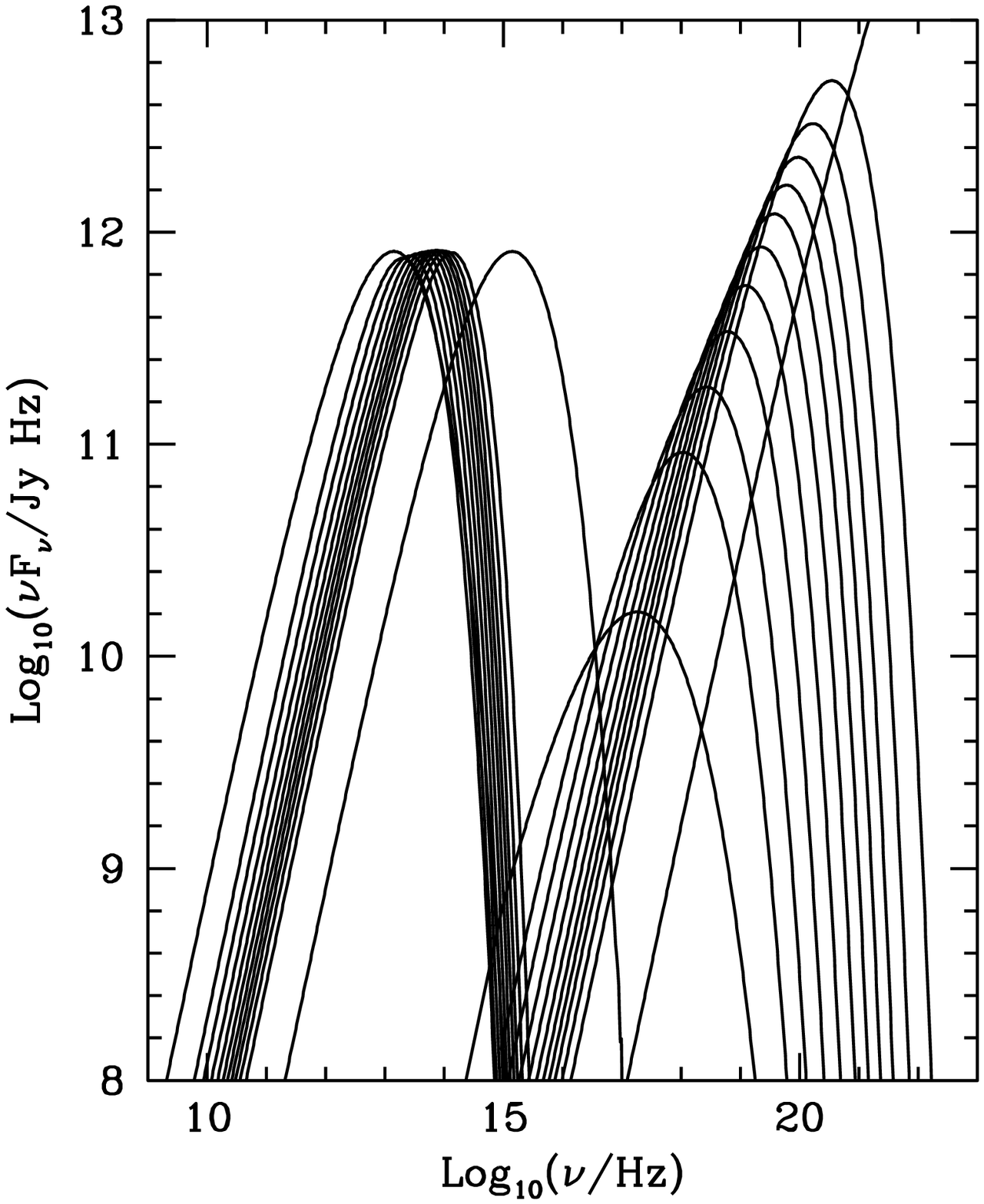}
\hspace{-0.6cm}
\includegraphics[height=8.4cm]{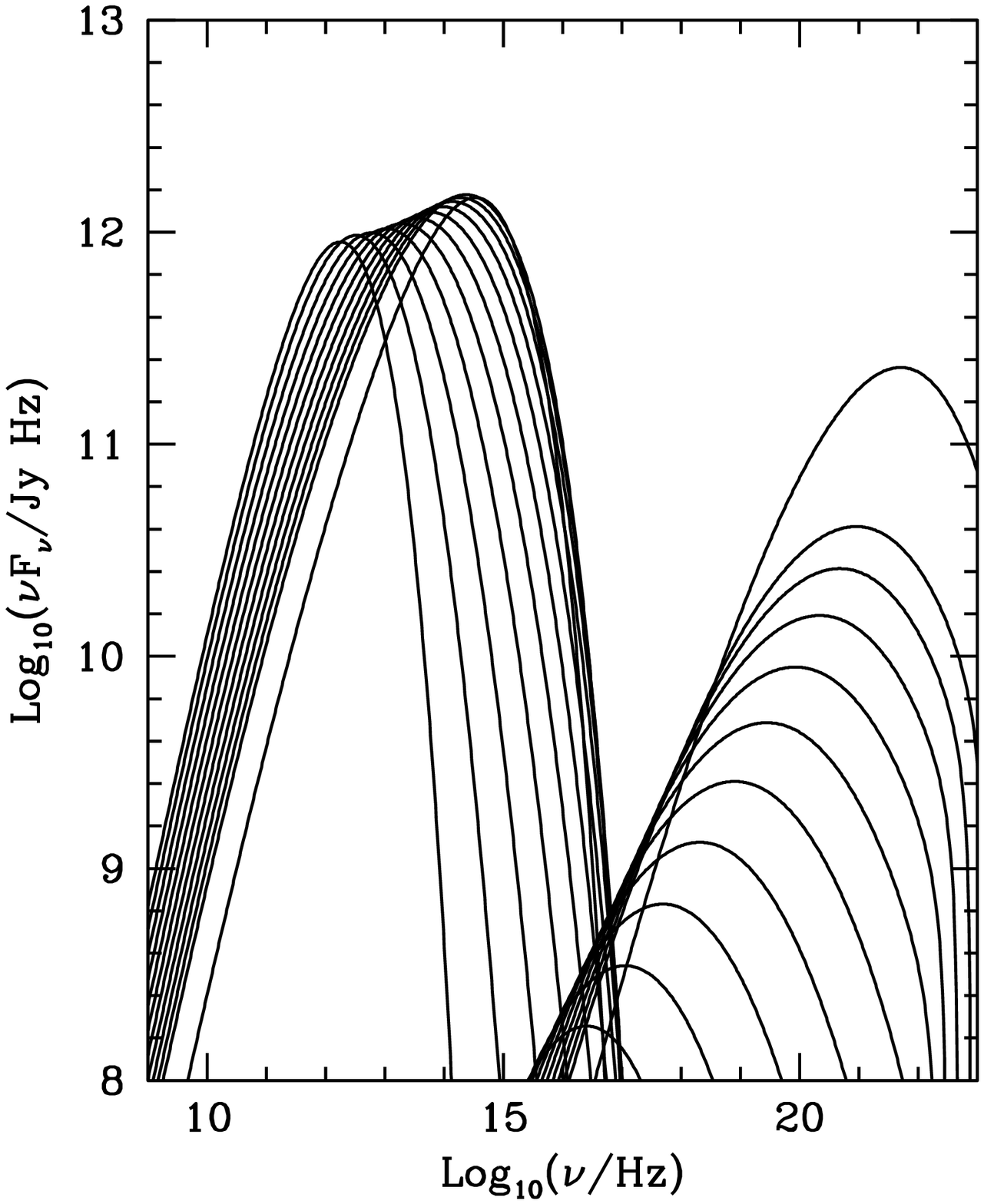}
\end{center}
\caption{
Evolution of the SSC spectrum during the cooling (left) and heating (right) phases shown in Figure \ref{fig8.ps}. Besides the 
initial and steady-state spectra, we show the spectral evolution for the cooling and heating phases when the correlation is 
produced. These spectra are indicated by the solid circles in Figure \ref{fig9.ps} and Figure \ref{fig11.ps}.
}
\label{fig10.ps}
\end{figure}

\begin{figure}[bht] 
\begin{center}
\includegraphics[height=8.4cm]{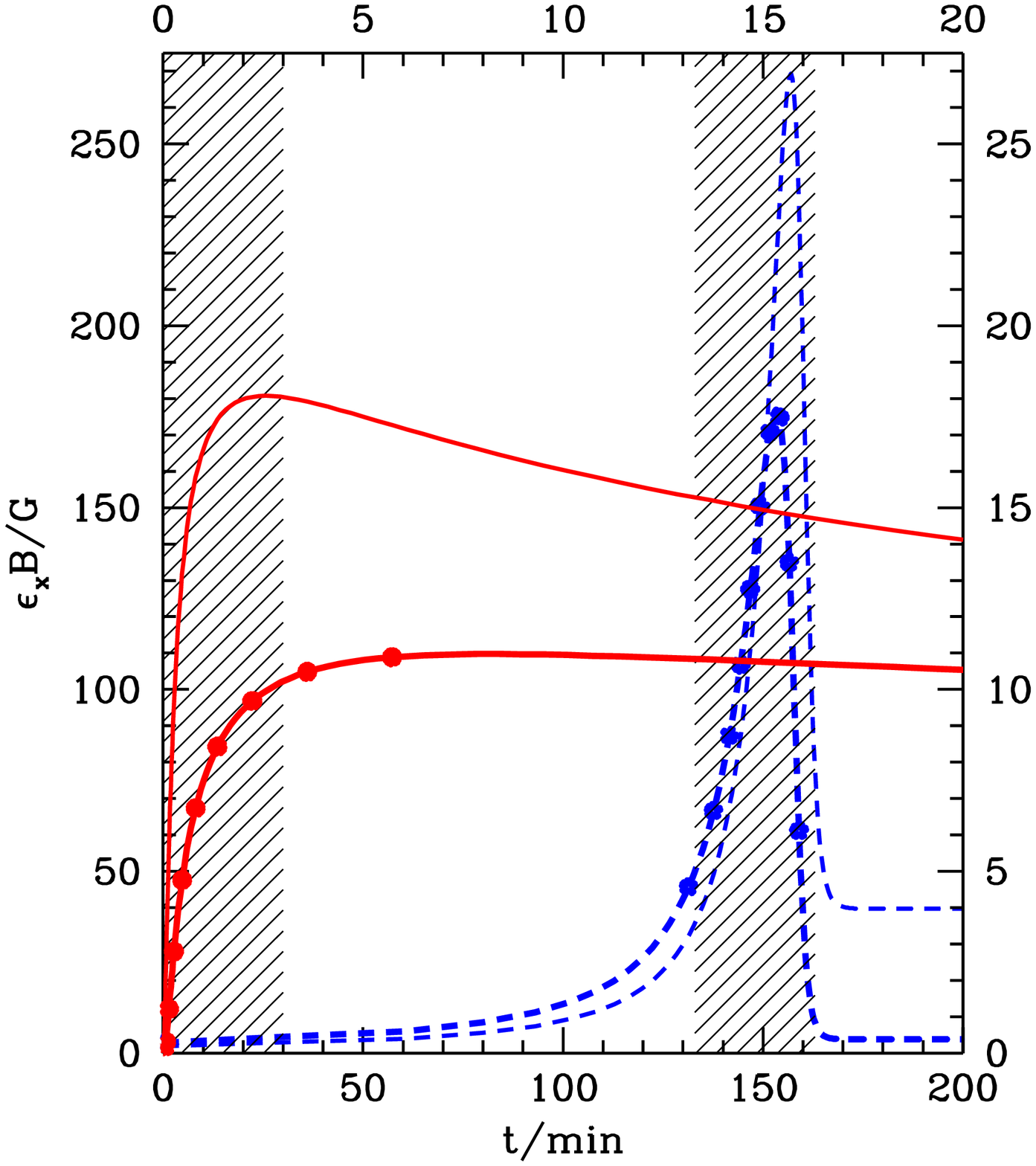}
\hspace{-0.6cm}
\includegraphics[height=8.4cm]{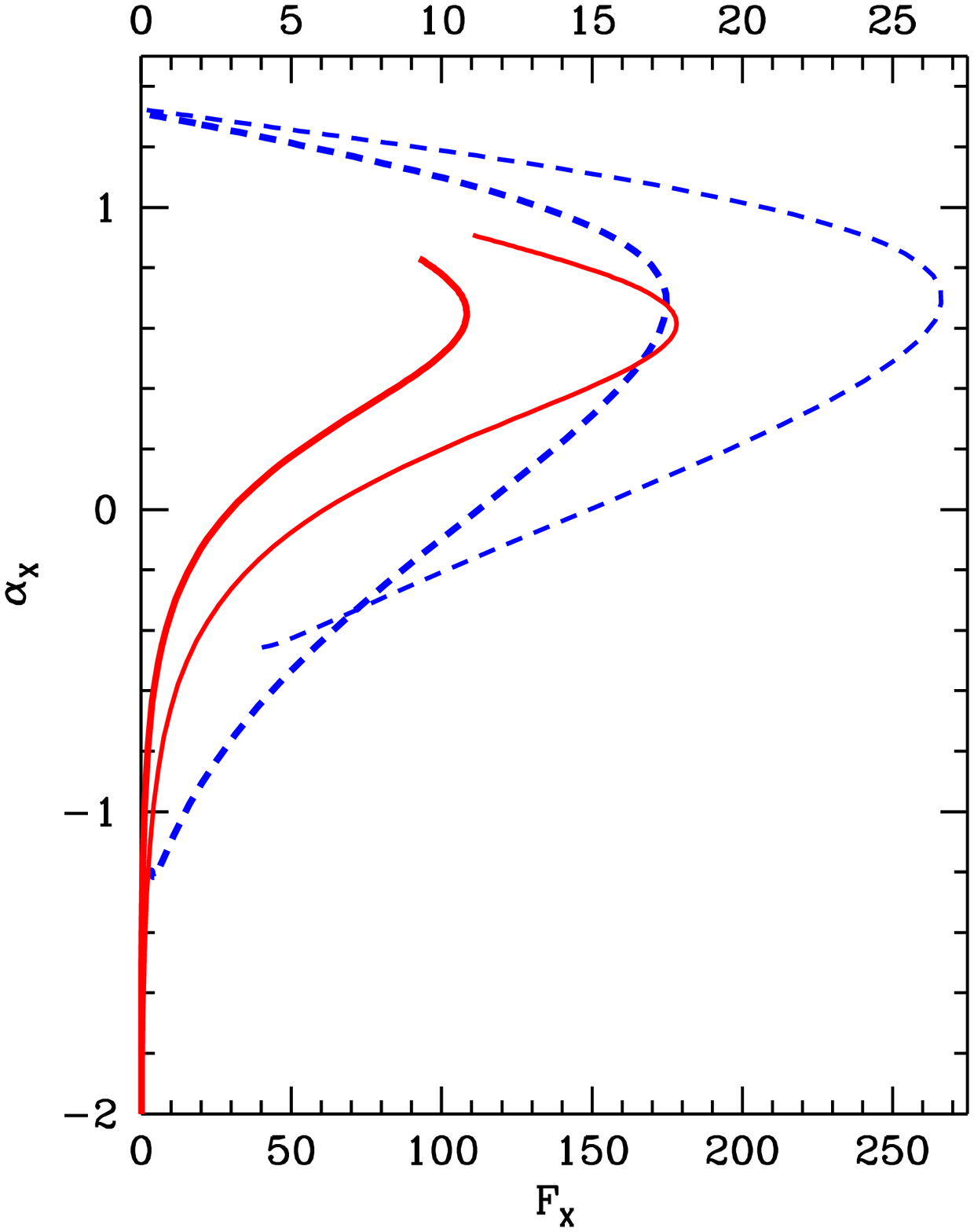}
\end{center}
\caption{
{\it Left:}
Light curve of the flux and spectral index at $0.5\times 10^{18}$ Hz (thin lines, corresponding to $\sim 2.1$ keV) and 
$2.0\times 10^{18}$ Hz (thick lines) during the cooling (dashed lines) and heating (solid lines) phases. The scales for the 
heating phase are indicated on the right and top axes.
{\it Right:} The corresponding correlation between the X-ray flux and spectral index. The scale of the heating phase is 
indicated on the top axis.  The solid circles correspond to the SSC spectra in Figure \ref{fig10.ps}.
}
\label{fig11.ps}
\end{figure}

The spectrum of the Comptonization component and its evolution should be quite different for the two phases, and may be used 
to distinguish between them. Figure \ref{fig10.ps} shows the evolution of the radiation spectrum during the cooling (left) and 
heating (right) phases. Besides indicating the initial and steady-state spectra, we highlight the spectral evolution when the 
correlation between NIR flux and spectral index are prominent, corresponding to the shaded periods in Figure \ref{fig9.ps} and Figure \ref{fig11.ps}, 
where the spectra are indicated by filled circles. For the model parameters chosen in \S \ref{vmag}, because the electron 
column depth $nR$ in the cooling phase is 50 times that in the heating phase while their synchrotron luminosities are 
comparable, much more high energy emission is produced via SSC in the cooling phase.

In analogy to equation (\ref{epss}), one can define a dimensionless SSC flux density spectrum
\begin{equation}
\epsilon_X(\nu)\equiv {6\sqrt{3}m_e^3 c^6 D^2F_X(\nu)\over \pi e^7 n^2 R^4 B}
 \simeq {\nu}\int_0^\infty 
\d\gamma {N(\gamma)\over \gamma^4}
\int_{\nu\over 4\gamma^2}^\infty 
\d \nu^\prime
{\epsilon_\nu(\nu^\prime)\over \nu^{\prime3}}
\left(2\nu\ln{\nu\over 4\gamma^2\nu^\prime}+\nu+4\gamma^2\nu^\prime-{\nu^2\over 2\gamma^2\nu^\prime}\right)\,.
\end{equation}
The left panel of Figure \ref{fig11.ps} shows the evolution of $B\epsilon_X$ at $0.5\times10^{18}$ (thin lines) and 
$2\times10^{18}$ Hz (thick lines) for the cooling (dashed lines) and heating (solid lines) phases. As expected, higher 
energy emission precedes (lags) lower energy emission slightly during the cooling (heating) phase because they are produced 
by more energetic electrons. However the difference in the peak times is within 10 minutes and may not be distinguished with 
the capacities of current instruments.  The time delay between the NIR and X-ray peaks is also within 30 minutes. We 
therefore expect a good correlation between the NIR and X-ray flux densities. The right panel shows the corresponding 
correlation between $\epsilon_X$ and the X-ray spectral index $\alpha_X$. This correlation gives a correlation between 
the X-ray flux density and spectral index, which can be tested with observations. 
Interestingly the X-ray flux density peaks near $\alpha_x\simeq 0.6-0.7$, which explains the hardness of X-ray spectra.

\subsection{Implications on the Source Structure}

The above studies clearly show that if the plasma producing the flares does not have complicated structure, the observed 
correlation between the spectral index and flux in the NIR band can only be explained by the cooling model with a magnetic 
field inversely proportional to the mean energy of the electrons.  This is especially true if  the total number of electrons 
involved does not change and the electron energy distribution is roughly uniform throughout the flare region. Other models 
either fit the observed correlation marginally or predict very weak dependence of the spectral index on the flux in the hard 
spectral states, which is inconsistent with observations. 

On the other hand, the flare region may have complicated source structure. The amount of electrons involved in producing the 
soft state emission can be quite different from that producing the hard state emission. If the electron distribution can be 
approximated as a Maxwellian spectrum, as is likely true, the observed correlation can then be used to determine the source 
structure. As shown in section \S\ \ref{conb}, $\alpha$ only depends on $B\gamma_0^2$, and $F_\nu$ also depends on $B n 
R^3$. For a given magnetic field, the observed correlation can be used to give a relation between the temperature $\gamma_0$ 
and the total number of electrons at such a temperature, which may shed light on the instabilities triggering the flares. 
One may also assume that the sum of the electron and magnetic field pressures is a constant in the flaring region. The 
observed correlation can also lead to a measurement of the amount of plasma at different temperatures. 

It is probably more natural to use MHD simulations to produce some flaring regions. The formalism developed in this paper can 
then be used to calculate the radiation spectrum. By comparing with the observed correlation, one can then determine what 
kind of active regions are likely responsible for the flares. It is clear that the SSC spectra are quite different for all 
these scenarios and simultaneous spectroscopic observations in the NIR and X-ray bands can be used to distinguish these 
models. Current observations do not justify such a comprehensive investigation, which is also beyond the scope of the paper. 

\section{Conclusion and Discussion}
\label{dis}

Flares from the direction of the supermassive black hole in the Galactic center are produced within a few Schwarzschild 
radii. Studying the details of the flare properties plays an important role in understanding the physics of black hole 
accretion. To infer the underlying physical processes with simultaneous spectroscopic observations over a broad frequency 
range, the plasma process of electron acceleration has to be addressed. Given the closeness of the flare variation timescale, 
the dynamical time at the last stable orbit of the black hole, and the synchrotron cooling time of electrons producing NIR 
emission during the flares, a time dependent study of the particle acceleration is necessary. Although there are already 
attempts to carry out this study analytically (Becker et al. 2006; Schlickeiser 1984), numerical solutions are required 
when specific events are considered.

In this paper we study the time dependent solutions of electron heating and cooling by a turbulent magnetic field. When the 
amplitudes of large scale magnetic fields are much smaller than the turbulent component, there are four basic model 
parameters, namely the magnetic field and its coherent length, the gas density and the initial distribution of electrons. 
Electrons gain energy from the plasma waves and loss energy via radiation, and the energy dependences 
of the heating and cooling rates are quite different. The evolution of the electron distribution is determined by these 
parameters.
We show that for the optically thin collisionless plasma in Sagittarius A*, synchrotron cooling dominates and 
relativistic Maxwellian distributions usually give a good description of the electron spectra. A formula is given to describe 
the dependence of the temperature evolution on the model parameters, which can be coupled with MHD simulations to study the 
radiation properties of the plasma.

We demonstrate that it is not trivial to produce the observed correlation between the spectral index and flux in the NIR 
band. Doppler effects and adiabatic processes suggested previously for the correlation (Gillessen et al. 2006; Yusef-Zadeh et 
al. 2006b) can not reproduce the observed results unless one has a very unusual electron spectrum. If the magnetic 
field does not change dramatically during the flare, only the cooling model of electrons heated instantaneously at the flare 
onset fits the observed correlation marginally. All other models predict very weak dependence of the spectral index on the 
flux density in the relatively hard spectral states.

We argue that the dynamical processes of electrons heating and cooling give the most reasonable explanation for the 
observations. To fit the correlation, the magnetic field needs to be anti-correlated with the electron temperature to reduce 
the flux density in the hard spectral states, bringing the above theoretical predictions more in line with the observations. 
It is not clear what mechanism may be responsible for such an anti-correlation. One possibility is that the dependence of the 
heating rate on the magnetic field is not as strong as that of the cooling rate. For example, the heating rate may be 
determined by the sound velocity, which can be much higher than the Alfv\'{e}n velocity. Moreover, the cooling due to inverse 
Comptonization will dominate when the electron temperature and column depth are very high, which is certainly the case in 
the early cooling phase\footnote{We note that the X-ray luminosity in the early cooling phase is higher than the synchrotron 
luminosity, suggesting a Compton catastrophe. This is due to the choice of the model parameters to embrace a large dynamical 
range. Including the cooling effect due to inverse Comptonization will make the electron temperature decrease rapidly to a 
value, where the Comptonization luminosity becomes lower than the synchrotron luminosity.}. All these effects can give rise 
to the desired anti-correlation.

The required anti-correlation between the magnetic field intensity and the electron mean energy suggests 
efficient energy exchanges between the field and electrons. While magnetic reconnection can cause transfer of magnetic field 
energy into electrons (corresponding to the heating phase), it is not clear what mechanism are responsible for the increase of 
magnetic field in the cooling phase. If the flare region is in pressure equilibrium with a relatively stable background, the 
magnetic field pressure may increase as the electron pressure decreases due to cooling. 

We note that the magnetic field pressure increases from 1.27 to $1.27 \times 10^4$ ergs cm$^{-3}$ and the electron 
pressure decreases from $6.1\times 10^5$ to $6.1\times 10^3$ ergs cm$^{-3}$ in the cooling phase of the flare discussed in 
\S\ \ref{vmag}. For the heating phase, the magnetic field pressure decreases from $3.2\times 10^3$ to 0.32 ergs cm$^{-3}$, 
while the electron pressure increases from 2.1 to $2.1\times 10^2$ ergs cm$^{-3}$. Although the pressures are far from 
equipartition in the initial and final states, the observed correlation is produced when the two are evolving toward an 
equipartition. These suggest that the flares are triggered by a process, which drives the energy densities of the magnetic 
field and electrons far from equipartition. The relaxation of the system toward energy equipartition produces the observed 
correlation. It is therefore critical to produce and/or identify features far from energy equipartition in MHD simulations to 
uncover the physical processes driving the flares. 

It is also possible that the flare region has complicated structure. The results of the electron spectral evolution 
presented here, which are also applicable to the relatively quiescent-state of the accretion flow,
then need to be coupled with MHD simulations to study the radiation characteristics of the accretion flow. 
The observed correlation between spectral index and flux suggests that, if the magnetic field does not change dramatically 
during the flare and through the flare region, more electrons are involved in producing the softer spectral 
state flux. Regions with small hot cores flashing in a extended lower temperature active envelope in MHD simulations may be 
identified with the flares.

\acknowledgments

This work was carried out under the auspices of the
National Nuclear Security Administration of the
U.S. Department of Energy at Los Alamos National
Laboratory under Contract No. DE-AC52-06NA25396.
This research was partially supported by NSF grant ATM-0312344, NASA grants NAG5-12111, NAG5 11918-1 (at Stanford),
and NSF grant PHY99-07949 (at KITP at UCSB). 
SL would like to thank Justin R. Pelzer for developing part of the code.

\end{document}